\documentclass[showpacs,pra,aps,superscriptaddress,twocolumn,floatfix]{revtex4}
\usepackage{graphicx,float,amsmath,amssymb}
\newfont{\ensmathquatorze}{msbm10 scaled 1400}
\newfont{\ensmathonze}{msbm10 scaled 1100}
\newfont{\ensmathdix}{msbm10}
\newfont{\ensmathneuf}{msbm10 scaled 833}
\newfont{\ensmathhuit}{msbm10 scaled 694}
\newfam\ensmathfam                        
\textfont\ensmathfam=\ensmathonze        
\scriptfont\ensmathfam=\ensmathdix       
\scriptscriptfont\ensmathfam=\ensmathhuit
\def\sss{\scriptscriptstyle}

\def\ep{{\mbox{\large e}}}

\begin{document}

\title{Anderson localization of a weakly interacting one dimensional Bose gas}

\author{T. Paul}
\affiliation{Laboratoire de Physique Th\'eorique
et Mod\`eles Statistiques, CNRS,
Universit\'e Paris Sud, UMR8626, 91405 Orsay Cedex, France}
\affiliation{Institut f\"ur Theoretische Physik, Universit\"at 
Heidelberg Philosophenweg 19, 69120 Heidelberg, Germany}
\author{M. Albert}
\affiliation{Laboratoire de Physique Th\'eorique
et Mod\`eles Statistiques, CNRS,
Universit\'e Paris Sud, UMR8626, 91405 Orsay Cedex, France}
\author{P. Schlagheck}
\affiliation{Institut f{\"u}r Theoretische Physik,
Universit{\"a}t Regensburg, 93040 Regensburg, Germany}
\affiliation{Mathematical Physics, Lund Insitute of Technology, PO Box 118,
22100 Lund, Sweden}
\author{P. Leboeuf}
\affiliation{Laboratoire de Physique Th\'eorique
et Mod\`eles Statistiques, CNRS,
Universit\'e Paris Sud, UMR8626, 91405 Orsay Cedex, France}
\author{N. Pavloff}
\affiliation{Laboratoire de Physique Th\'eorique
et Mod\`eles Statistiques, CNRS,
Universit\'e Paris Sud, UMR8626, 91405 Orsay Cedex, France}
\begin{abstract}
We consider the phase coherent transport of a quasi one-dimensional beam of
Bose-Einstein condensed particles through a disordered potential of length
$L$.  Among the possible different types of flow we identified [T. Paul, P.
  Schlagheck, P. Leboeuf and N. Pavloff, Phys. Rev. Lett. {\bf 98}, 210602
  (2007)], we focus here on the supersonic stationary regime where Anderson
localization exists.  We generalize the diffusion formalism of
Dorokhov-Mello-Pereyra-Kumar to include interaction effects. It is
shown that interactions modify the localization length and also introduce a
length scale $L^*$ for the disordered region, above which most of the
realizations of the random potential lead to time dependent flows. A
Fokker-Planck equation for the probability density of the transmission
coefficient that takes this new effect into account is introduced and
solved. The theoretical predictions are verified numerically for different
types of disordered potentials.  Experimental scenarios for observing our
predictions are discussed.
\end{abstract}

\maketitle


\section{Introduction}
The absence of diffusion of waves in disordered media was predicted by
Anderson 50 years ago \cite{And58}. Originally proposed in the context of
electronic transport in disordered crystals, it has since been observed for
different types of waves, including light and sound. Recently, direct
observations of the Anderson localization by disorder \cite{Bil08} and of a
localization transition by quasiperiodic potentials \cite{Roa08} of quasi
one-dimensional (1D) matter waves of ultracold atoms were reported. These
experiments pave the way to the observation of new phenomena and shed new
light on long standing problems, amongst which the question of possible
Anderson localization in presence of interactions.

In the present paper we consider the case of an atomic vapor described as a
weakly interacting Bose gas in the presence of a weak disorder (what is meant
by ``weak'' here will be made quantitative in Sec. \ref{model}).  In this
configuration it has been shown theoretically in Refs.~\cite{Hua92,Gio94} and
supported by numerical simulations in Ref.~\cite{weakdis2} that a small amount
of disorder does not drastically alter the {\sl equilibrium} properties of the
system, but merely decreases the condensate and superfluid fractions.
Furthermore, even in the 1D limit considered in the present work, it has been
experimentally demonstrated in Refs.~\cite{Cle08,Che08} that one can observe
global phase coherence in the presence of disorder and remain far from, say,
the Bose glass phase originally proposed by Giamarchi and Schulz and Fisher
{\it et al.} \cite{GiaSch,Fish}.

\begin{figure}
\includegraphics*[width=0.95\columnwidth]{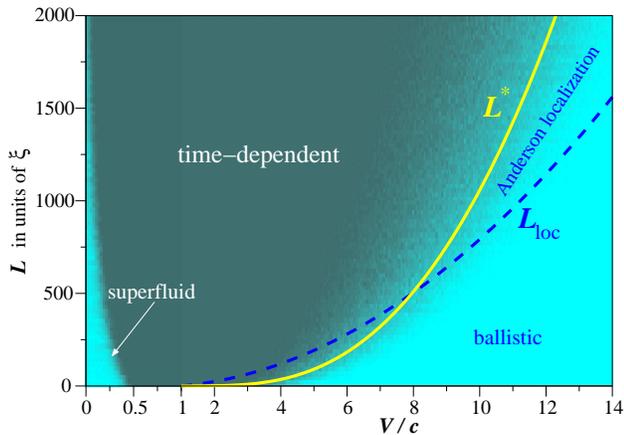}
\caption{ (Color online) Transport of a quasi 1D BEC with velocity $V$ through
  a disordered potential $U_\delta$ consisting in a series of uncorrelated
  delta peaks extending over a domain of size $L$ [cf.  Eq.~(\ref{e2a}) and
    the discussion in Sec. \ref{sec-lstar}].  Dark region: time dependent
  flow; light gray (light blue online) regions: stationary flow. In the
  supersonic case, the yellow solid line corresponds to the threshold $L^*$
  between these two domains as determined from Eq. (\ref{tstar-impl}). The
  blue dashed line is the localization length $L_{\rm loc}$ (\ref{exp1}). The
  supersonic region below $L_{\rm loc}$ denoted as ``ballistic'' corresponds
  to the region where the perturbation theory of Sec. \ref{pts} applies. Note
  the enlarged scale for $V/c\in[0,1]$. \label{fig1}}
\end{figure}

Here we are interested in {\sl transport} properties.  Specifically, we study
a quasi 1D, weakly interacting Bose Einstein condensate (BEC), propagating
through a disordered potential.  In this context, localization has been
theoretically studied mainly for effective {\it attractive} interactions (see,
e.g., \cite{Gre92} and references therein), with less attention on the {\it
  repulsive} case we consider here (see, however,
Refs.~\cite{Bil05,Pau05}). In the absence of an external potential,
(repulsive) interactions make the system superfluid and introduce a new
characteristic speed in the system, the speed of sound $c$. As mentioned
above, when the speed $V$ of the BEC relative to the external potential tends
to zero, the addition of a weak random potential preserves superfluidity,
although with a reduced superfluid fraction. What happens as $V$ increases ?
This question was investigated in a previous publication \cite{Pau07}, where
the disordered potential, of length $L$, was modeled by a series of randomly
located delta peaks. For small velocities $V/c \ll 1$, perturbation theory
shows that the superfluidity is preserved, e.g., the flow is dissipationless
and with a perfect transmission. In contrast, in the high speed limit $V/c \gg
1$, where the kinetic energy dominates over the interaction energy, the
transport properties of the BEC are deeply altered, and tend to those of the
non-interacting gas, displaying an exponential damping of the transmission
with length $L$, a behavior characteristic of the strong Anderson
localization. Thus, two limiting cases of stationary flow have been identified
\cite{Pau07}, with contrasting transport properties: superfluidity in the deep
subsonic regime, and Anderson localization in the deep supersonic one. In
between, in the region $V\sim c$ where both interaction and kinetic energies
are important, it was shown that stationary scattering solutions do not exist:
one reaches a regime of time-dependent flows with more or less (depending on
the speed) complex density excitations.  The range of speeds around $c$ where
this phenomenon is observed increases as the length $L$ increases. The
different types of existing flows are summarized in Fig.~\ref{fig1}.

In the present study we concentrate on the supersonic stationary region of the
phase diagram [gray (light blue online) $V/c>1$ region in Fig.~\ref{fig1}]. In
this domain we provide analytical and numerical evidence of Anderson
localization in the presence of interaction for different types of
disorder. We compute analytically the interaction-dependent localization
length as well as the corresponding distribution of transmission
coefficients. We also explain the disappearance of the supersonic stationary
flow observed at a given velocity for increasing length of the disordered
sample. This onset of time dependence is an important qualitative effect
revealed by our study. We show that it is directly connected to interaction
effects and provide an analytical estimate of the length $L^*$ of the
disordered region above which most of the realizations of the random potential
lead to time dependent flows (see Fig.  \ref{fig1}).

The paper is organized as follows. In Sec. \ref{model} we present the model
and identify its range of validity. In Sec. \ref{secIII} we take some time to
properly define the transmission coefficient of a Bose-condensed beam over an
obstacle. In \ref{secIV} we introduce the different types of disordered
potentials studied in the present work. In Sec. \ref{ssr} we present
analytical and numerical results showing that Anderson localization is indeed
possible in the supersonic regime. We consider the three possible supersonic
regimes: perturbative (Sec. \ref{pts}), Anderson localized
(Sec. \ref{non-pert}) and onset of time dependence (Sec. \ref{sec-lstar}). In
Sec. \ref{experiment} we discuss experimental strategies and possible
signatures for the observation of Anderson localization in an interacting
Bose-Einstein condensate. Finally we present our conclusions in
Sec. \ref{conclusion}.  Some technical points are given in the appendixes. In
Appendix \ref{app1} we derive the probability distribution of transmission in
a special case (perturbative regime and correlated Gaussian potential).  In
Appendix \ref{app2} we present the derivation of the Fokker-Plank equation
(\ref{dmpk}) for the distribution of the transmission coefficients.

\section{Model}\label{model}

We study here the transport properties of a quasi-one-dimensional (1D)
Bose-Einstein condensate formed of particles of mass $m$, experiencing a
repulsive effective interaction (characterized by the 3D s-wave scattering
length $a>0$), in the presence of an obstacle represented by the external
potential $U$. The potential is not necessarily disordered at this point, the
only restriction we impose throughout the present work is that it should have
a finite extent, i.e., $U(x)\to 0$ when $x\to\pm\infty$.  The configuration we
consider corresponds to the ``1D mean field regime'' \cite{Men02} (see also
the discussion in Ref.~\cite{Pet00}), where the system is described by a 1D
order parameter $\psi(x,t)$ depending on a single spatial variable: the
coordinate $x$ along the direction of propagation.  $\psi(x,t)$ obeys the
nonlinear Schr\"odinger equation
\begin{equation}\label{e1}
{\rm i}\,\hbar\,\frac{\partial\psi}{\partial t}=-\frac{\hbar^2}{2
	m}\frac{\partial^2\psi}{\partial x^2} + \left[
U(x-Vt)+g\,|\psi|^2-\mu\right]\psi\; .
\end{equation}
In all the present work we choose to work in the ``laboratory frame'' where
the condensate is initially at rest. Eq. (\ref{e1}) describes its 1D dynamics
in the presence of an obstacle moving at constant velocity $V$ in this frame,
which corresponds to the experimental situation where an obstacle is swept
through a condensate initially at rest, see
e.g. Refs.~\cite{Ram99,Ono00,Eng07}.  On the theoretical side, one should
imagine that, from an initial static configuration where the condensate is at
rest with $U\equiv 0$, the potential intensity and speed have been slowly
ramped up to a point where the condensate dynamics is described by
Eq. (\ref{e1}). We choose $V>0$, this corresponds to a potential moving from
left to right in the laboratory frame.

The reduction of the motion of the condensate to a single spatial dimension is
typically achieved through a transverse harmonic confining potential of
pulsation $\omega_\perp$.  We choose a normalization such that
$n(x,t)=|\psi(x,t)|^2$ is the linear density of the condensate. In this case,
the interaction amongst particles results in Eq. (\ref{e1}) in the nonlinear
term $g|\psi|^2$, with $g=2\,\hbar\,\omega_\perp\, a$
\cite{Ols98,Jac98,Leb01}.

In the stationary regime, where the flow is time--independent in the frame
moving with the potential, $\psi$ depends on $x$ and $t$ only through the
variable $X=x-Vt$. The appropriate boundary condition is
$\psi(X\to-\infty)=\sqrt{n_0}$ (where $n_0$ is a constant) (see \cite{Leb01}
and the discussion in Section \ref{stationary_regime} below). The condensate
is then characterized by a chemical potential $\mu=g\, n_0$, a speed of
sound $c=(g\, n_0/m)^{1/2}$ and a healing length $\xi=\hbar/(m\,c)$.
  
It is customary to characterize the transverse confinement via the ``harmonic
oscillator length'' $a_\perp=(\hbar/m\omega_\perp)^{1/2}$. With $n_1$
denoting a typical order of magnitude of $n(x,t)$, the 1D mean field
regime in which Eq. (\ref{e1}) is valid
corresponds to a density range such that
\begin{equation}\label{s1.0}
(a/a_\perp)^2\ll n_1\,a \ll 1\; .
\end{equation}
In this domain the wave function of the condensate can be factorized in a
transverse and a longitudinal part \cite{Jac98,Ols98,Leb01}. The transverse
wave function is Gaussian (this is ensured by the condition $n_1\,a \ll 1$),
the longitudinal one is of the form $\psi(x,t)\exp\{-{\rm i}\,\mu t/\hbar\}$
and $\psi(x,t)$ satisfies Eq.~(\ref{e1}) \cite{Jac98,Leb01}. The left-hand
side (l.h.s.) inequality in Eq.~(\ref{s1.0}) prevents the system to enter in
the Tonks-Girardeau regime.  More precisely, a general analysis of 1D Bose gas
shows that at zero temperature no BEC is possible \cite{SP91}. This results in
a algebraic decrease of the one body density matrix monitored by phase
fluctuations occurring over a phase-coherence length $L_\phi=\xi \exp
\{\pi\,a_\perp (n_1/2\,a)^{1/2}\}$ \cite{Sch77,Petrov}. Hence the results
obtained using Eq.~(\ref{e1}) are valid if they describe structures with a
characteristic length scale smaller than $L_\phi$. The l.h.s. inequality in
Eq. (\ref{s1.0}) ensures that $L_\phi$ is exponentially large compared to the
healing length.  If one considers, for instance, $^{87}$Rb or $^{23}$Na atoms
in a guide with a transverse confinement characterized by
$\omega_\perp=2\pi\times 500$ Hz, the ratio $a/a_\perp$ is roughly of order
$10^{-2}$ and restriction (\ref{s1.0}) still allows the density to vary
over four orders of magnitude.

Even if the mean field approach is legitimate in 1D, the effects of disorder
have to be taken into account with some care. It may well be that the
introduction of a disordered potential $U$ in Eq.~(\ref{e1}) modifies the
properties of the ground state. This is indeed the case as shown in
Refs.~\cite{Hua92,Gio94}~: a disordered potential decreases the condensate and
the superfluid fraction, but the effects are weak provided the intensity of
the disorder remains weak (see Ref.~\cite{Lop02} for an extension to finite
temperature). More precisely, in the case of a disorder formed by randomly
spaced delta impurities with density $n_{\delta}$ (see Sec.~\ref{delta}) one
can show \cite{Ast04,Pau07} that, in the dilute impurity limit, at $V=0$ the
non-superfluid fraction (normal part) is proportional to $n_{\delta}\xi
(\xi/b)^2$ [the notations are those of Eq.~(\ref{e2a})] and thus remains small
provided the dimensionless coefficient $(\xi/b)$ is small (weak disorder
limit). At finite $V$, the normal fraction is multiplied by a factor
$[1-(V/c)^2]^{-3/2}$ (see Ref.~\cite{Pau07}), which diverges when $V=c$. One
thus expects the mean field approach to fail near the region $V\simeq c$ of
Fig. \ref{fig1}.  This is supported by the numerical results presented in
\cite{Ern09}. Hence, in the center of the time-dependent region of
Fig. \ref{fig1} we cannot trust the results obtained from
Eq.~(\ref{e1}). However, far from this region, the 1D mean field approach is
expected to be valid even in presence of (weak) disorder, as experimentally
demonstrated in Refs.~\cite{Cle08,Che08}.

\section{Definition of the transmission}\label{secIII}

In the present work we characterize the localization properties of the
condensate in the random potential by studying the transmission
coefficient. Eq.~(\ref{e1}) being non linear, the definition of transmission
and reflection coefficients needs to be treated with special care. This is the
purpose of the present section where we first define the stationary regime
(Section \ref{stationary_regime}) and then the transmission coefficient within
this regime (Section \ref{transmis}).

\subsection{Stationary regime}\label{stationary_regime}

It is customary to perform a Madelung transformation and to write $\psi(x,t) =
\sqrt{n(x,t)} \exp\{ {\rm i}\, S(x,t)\}$ where $n(x,t)$ is the density and
$\hbar \, \partial_x S/m=v(x,t)$ the local velocity. From (\ref{e1}) one can
check that they verify the continuity equation
\begin{equation}\label{e2}
\partial_t n +\partial_x(n v)=0 \; .
\end{equation}
The stationary regime is defined as the regime where the system is at rest in
the frame moving with the obstacle. In this case, in the laboratory frame
$\psi$, $S$, $n$ and $v$ are time dependent, but they depend on $x$ and $t$
only through the variable $X=x-Vt$. It is then possible to get a first
integral of (\ref{e2}) under the form
\begin{equation}\label{e3}
n(X)\left(
\frac{\hbar}{m} \frac{{\rm d}S}{{\rm d}X} -V\right)=C^{\rm st} \; .
\end{equation}

In the case of subsonic ($V<c$) and stationary motion, the flow is superfluid
and the order parameter is only affected in the vicinity of the obstacle, with
$n(X\to\pm \infty)=n_0$ and $v(X\to\pm \infty)=0$ \cite{Hak97,Leb01}.

\begin{figure}
\begin{center}
\begin{picture}(9,5.5)
\put(1,0){\includegraphics[width=8cm]{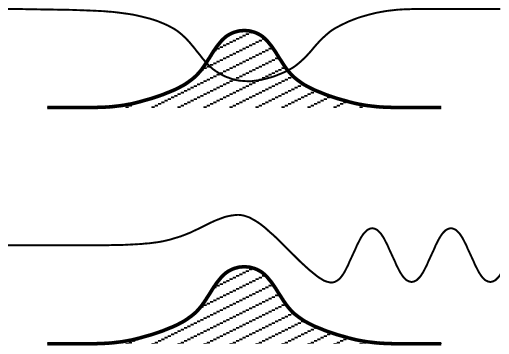}}
\put(0.5,3.){\large\sffamily downstream}
\put(6.7,3.){\large\sffamily upstream}
\put(1,2.7){\vector(1,0){7.5}}
\put(6.6,2.3){$X=x-Vt$}
\put(1.4,1.7){$n(X)$}
\put(1.4,5.3){$n(X)$}
\put(3.2,1.){$U(X)$}
\put(3.8,5.){$U(X)$}
\put(0,1.5){$n_0$}
\put(0.4,1.55){\vector(1,0){0.5}}
\put(0,5.03){$n_0$}
\put(0.4,5.08){\vector(1,0){0.5}}
\end{picture}
\end{center}
\caption{Schematic representation of the typical density profiles.  The upper
  plot corresponds to a subsonic stationary profile, while the lower one
  corresponds to a supersonic stationary profile. The potential moves from
  left to right, and the upstream (downstream) region thus corresponds to the
  region $X\to+\infty$ ($X\to-\infty$). In both plots the potential is
  represented by a thick solid line (hatched down to zero) and the density
  profile is represented by a thin solid line.\label{fig2}}
\end{figure}

For $V>c$, a regime of stationary flow also exists but in this case the
obstacle induces density oscillations with a pattern stationary in its rest
frame \cite{Leb01}. This means that in the laboratory frame the phase velocity 
of these waves is identical to the velocity
$V$ of the obstacle.  On the other hand, the energy transferred from the
obstacle to the fluid propagates with the group velocity, which in the case of
Bogoliubov excitations is greater than the phase velocity, i.e. -- as just
argued -- than $V$. As a consequence, radiation conditions require that the
wake is always located ahead of the obstacle, i.e., upstream, with no
long-range perturbation of the fluid on the downstream side
\cite{Leb01,Lamb}. This means that in this case the flow far in the downstream
region remains unperturbed, with $n(X\to -\infty)=n_0$ and $v(X\to
-\infty)=0$. The two possible stationary configurations (subsonic and
supersonic) are represented in Fig.~\ref{fig2}.

Hence, in any stationary configuration (subsonic or supersonic), the above
reasoning fixes the integration constant in the right hand side (r.h.s.) of
Eq.~(\ref{e3}) to its value at $X\to -\infty$, i.e., $-n_0 V$.

In the stationary regime one gets from Eqs. (\ref{e1}) and (\ref{e3})
\begin{equation}\label{e4}
U(X)\, \frac{{\rm d}A^2}{{\rm d}X} =
\frac{{\rm d}}{{\rm d}X}\left\{ 
\frac{\hbar^2}{2 m}\left(\frac{{\rm d}A}{{\rm d}X}\right)^2+
W(A)\right\} \; ,
\end{equation}
\noindent where $A(X)=\sqrt{n(X)/n_0}$ and
\begin{equation}
W(A)=\frac{m}{2} (A^2-1) 
\left[c^2+V^2-c^2 A^2 -\frac{V^2}{A^2}\right]
\; .
\end{equation}

\subsection{Transmission coefficient}\label{transmis}

In this section we restrict the analysis to the stationary regime of section
\ref{stationary_regime}, and define the transmission of the condensate through
the obstacle represented by a potential $U$ (not necessarily disordered)
verifying $U(|x|\to\infty)=0$. 

As the wave equation (\ref{e1}) is nonlinear, one cannot, in general, properly
define reflection and transmission coefficients, since it is generally not
possible to disentangle incoming and reflected waves in the nonlinear flow
upstream the obstacle.  However, following a procedure devised in
Ref.~\cite{Leb03} (see also \cite{Pau07b}), we will show that one can define a
transmission and a reflection coefficient in the limit of small nonlinearity
as well as in the limit of weak reflection and arbitrary nonlinearity.

\begin{figure}
\begin{center}
\includegraphics[width=0.95\columnwidth]{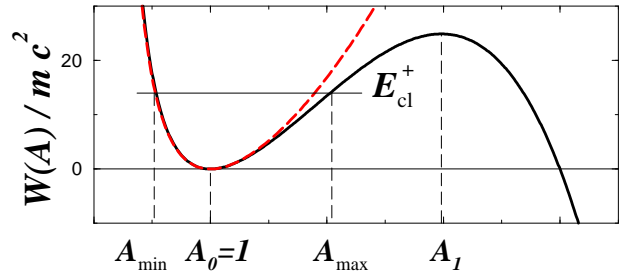} 
\end{center}
\caption{(Color online) $W$ as a function of $A=|\psi|/\sqrt{n_0}$ (drawn for
  $V/c=4$). The fictitious particle has a classical energy $E_{\rm cl}^+$ when
  $X\to +\infty$. The (red) dashed line corresponds to an approximation of
  $W(A)$ by $\hbar^2\kappa^2(A-1/A)^2/(2 m)$, obtained by keeping only the
  first term in expansion (\ref{t3a}).\label{fig3}}
\end{figure}

Outside the scattering region, $U(X)=0$ and one can get a first
integral of Eq.~(\ref{e4}) under the form
\begin{equation}\label{t1} 
\frac{\hbar^2}{2 m}\left(\frac{{\rm d}A}{{\rm d}X}\right)^2+
W(A)=E^{\pm}_{\rm cl} \; , \quad\mbox{when}\quad
X\to\pm\infty \; ,
\end{equation}
which defines the ``free'' asymptotic density profiles.  $E^{\pm}_{\rm cl}$ in
Eq.~(\ref{t1}) are integration constants. The boundary condition discussed in
the previous section imposes $A=1$ and ${\rm d}A/{\rm d}X=0$ when $X \to
-\infty$. This fixes the value $E_{\rm cl}^-=0$. The value of $E_{\rm cl}^+$
at $+\infty$ has to be determined by the integration of Eq.~(\ref{e1})
(cf. Ref.~\cite{Leb01}).  Eq.~(\ref{t1}) expresses the energy conservation for
a fictitious classical particle with ``mass'' $\hbar^2/m$, ``position'' $A$
and ``time'' $X$, evolving in a potential $W$ (whose typical shape is
displayed in Fig.~\ref{fig3}).  This type of analysis is common in the study
of nonlinear equations such as Eq. (\ref{e1}), see e.g., the review
\cite{Ivl84} (the first time we found it used is in Ref. \cite{Lan67}). It is
employed here as a convenient tool for getting intuition about the behavior of
the solution of the Gross-Pitaevskii equation (see below).

From now on we restrict to the supersonic stationary regime where an imperfect
transmission occurs (in the subsonic stationary regime one has perfect
transmission). In this case the fictitious particle is initially (i.e., when
$X\to-\infty$) at rest at the bottom of the potential $W$ with $E_{\rm
  cl}^-=0$.  The behavior of $A$ for $X\to+\infty$ depends on the value of
$E^{+}_{\rm cl}$. A stationary solution exists only if $A(X\to+\infty)$
remains bounded, i.e., if $E^{+}_{\rm cl}<W(A_1)$ (where $A_1$ corresponds to
the local maximum of $W$, see Fig. \ref{fig3}). In this case, the asymptotic
behavior of $A(X)$ corresponds to oscillations between the values $A_{\rm
  min}$ and $A_{\rm max}$ defined in Fig. 3.

For future references we note that $W(A)$ is zero when $A=A_0=1$ and when
$A=V/c$, and that the derivative ${\rm d}W/{\rm d}A$ is zero when $A=A_0=1$
and when $A=A_1$, with
\begin{equation}\label{t1a}
A_1=\frac{V}{2c} \left(1+\sqrt{1+\frac{8 c^2}{V^2}}\right)^{1/2}\; .
\end{equation}
At large velocity, when $V\gg c$, one has
\begin{equation}\label{t1b}
A_1=\frac{V}{\sqrt{2} \, c}+{\cal O}\left(\frac{1}{V/c}\right)
\; ,
\end{equation}
and
\begin{equation}\label{t1c}
W(A_1) = \frac{m V^4}{8\,c^2} 
+{\cal O}\left(\frac{V^2}{c^2}\right)\; .
\end{equation}

Writing $A^2(X)=\rho(X)=1+\delta\rho(X)$ we now argue that, following
Ref.~\cite{Leb03}, one can write a perturbative
version of (\ref{t1}) in a limit where
\begin{equation}\label{t2}
|\delta\rho(X)| \ll
\left|\frac{V^2}{c^2}-1\right| \; .
\end{equation}
We emphasize that restriction (\ref{t2}) corresponds to $|\delta\rho| \ll
1$ (i.e., small oscillations) only when $V$ is of order of $c$ or
smaller. The approach developed below is however able to tackle large relative
density oscillations ($|\delta\rho| \gg 1$) at large velocities ($V\gg c$)
\cite{rem}. In this sense it will allow us to penetrate in the non
perturbative regime where the upstream density oscillations are large and the
transmission is low.

Using the variable $\rho$, we write Eq.~(\ref{t1}) in the upstream region
($X\to +\infty$) as
\begin{equation}\label{t3}
\frac{\hbar^2}{2 m}\left(\frac{{\rm d}\rho}{{\rm d}X}\right)^2+
8 F(\rho)=8\rho E_{\rm cl}^+ \; ,
\end{equation}
where $F(\rho)=\rho\, W(A=\sqrt{\rho})$. A simple limited expansion around
$\rho=1$ yields
\begin{equation}\label{t3a}
F(\rho)\simeq
\frac{\hbar^2 \kappa^2}{2 m}(\delta\rho)^2+\frac{m c^2}{2}(\delta\rho)^3
+\cdots
\end{equation}
where $\delta \rho(X)=\rho(X) - 1$ and
\begin{equation}\label{pt4}
\kappa=\frac{m}{\hbar}\left|V^2-c^2\right|^{1/2}
\; , \quad \mbox{and}\quad
\left|\frac{V^2}{c^2}-1\right| = \kappa^2\xi^2\; .
\end{equation}
The second term in the r.h.s. of expansion (\ref{t3a}) is small compared to
the first one precisely in the limit (\ref{t2}). In the following we restrict
to this regime and neglect the second term of the r.h.s. of (\ref{t3a}). This
corresponds to approximating the exact $W(A)$ by the (red online) dashed line
in Fig. \ref{fig3} and to write Eq.~(\ref{t3}) under the form
\begin{equation}\label{t4}
\left(\frac{{\rm d}\delta\rho}{{\rm d}X}\right)^2+4\kappa^2 (\delta\rho)^2
= 16 \kappa^2\lambda (1+\delta \rho) \; ,
\end{equation}
where the dimensionless parameter $\lambda$ is defined by
\begin{equation}\label{t5}
\lambda=\frac{m \, E^+_{\rm cl}}{2\, \hbar^2 \kappa^2}\; .
\end{equation}
The solution of Eq.~(\ref{t4}) is
\begin{equation}\label{t6}
\frac{n(X)}{n_0}=\rho(X)=1+2\,\lambda+2\,\Lambda
\cos(2\kappa X+\theta) \; ,
\end{equation}
where
\begin{equation}\label{t6b}
\Lambda=\sqrt{\lambda^2+\lambda} \; ,
\end{equation}
and $\theta$ is a phase factor. We recall that Eq.~(\ref{t6}) describes the
density oscillations in the upstream region $X\to +\infty$.  These
oscillations can be described as the sum of incident and reflected waves
($\psi_{\rm inc}$ and $\psi_{\rm ref}$) of the form
\begin{eqnarray}\label{t6c}
\psi_{\rm  inc}(X) & = & 
\sqrt{n_0(1+\lambda)}\, \exp(-i\kappa X) \; , \nonumber \\
\psi_{\rm ref}(X) & = & 
\sqrt{n_0\lambda} \, \exp(i\kappa X+i\theta) \; . 
\end{eqnarray}
This analysis allows one to determine the reflection and the transmission
coefficients as
\begin{equation}\label{t7}
R=\frac{|\psi_{\rm ref}|^2}{|\psi_{\rm  inc}|^2}
=\frac{\lambda}{1+\lambda} \; , \quad T=1-R=\frac{1}{1+\lambda} \; .
\end{equation}
Of course the sum of the incident $\psi_{\rm inc}$ and the reflected $\psi_{\rm
ref}$ waves (\ref{t6c}) is an approximate solution of the nonlinear
Sch\"odinger equation (\ref{e1}) which is only valid in regime
(\ref{t2}), i.e., in the regime of arbitrary interaction and small
transmission ($\lambda\ll 1$), or in the regime of arbitrary transmission and
small interaction ($V\gg c$).

\section{Different types of disorder}\label{secIV}

Up to this point we presented a theory valid for any potential of finite
extent. From now on we concentrate on the particular case of random 
potentials. We denote $U(x)$ an arbitrary random potential, and use
a subscript when dealing with one of the particular cases defined below. 

\subsection{Potential formed by a series of $\delta$ peaks}\label{delta}

The first potential of interest, analyzed in Ref.~\cite{Pau07}, is a 
series of $N$ randomly located identical delta peaks of the form~:
\begin{equation}\label{e2a}
U_{\delta}(x)=\frac{\hbar^2}{m\,b} \sum_{i=1}^{N}\delta(x-x_i) \; .
\end{equation}
The intensity of the peaks is measured by the (non random) positive quantity
$b$. The scatterers have random uncorrelated positions $0=x_1\le x_2\le
x_3...$, with mean density $n_\delta$ and average separation
$l_\delta=1/n_\delta$.  Hence the potential extends over a mean length
$L=(N-1) l_\delta$.

Denoting henceforth the disorder average by $\langle ..\rangle$, for $x$ and
$x'$ inside the disordered region one gets the mean value
\begin{equation}\label{e2b0}
\langle U_{\delta}(x) \rangle =\frac{\hbar^2 n_\delta}{m\, b} \; ,
\end{equation}
and the irreducible two-point correlation function
\begin{equation}\label{e2b}
\langle U_{\delta}(x)U_{\delta}(x')\rangle-\langle U_{\delta}\rangle^2
=\sigma\,(\hbar^2/m)^2
\,\delta(x-x')\; ,
\end{equation}
where
\begin{equation}\label{e2c}
\sigma=\frac{n_\delta}{b^2} \; .
\end{equation}

\subsection{Correlated Gaussian potential}\label{gaussian}

Another commonly used model of disorder is provided by Gaussian random
processes with zero average. We consider here potentials which are non zero
only over a region of finite extent (with typical size $L$) and generate them
in the following way (see, e.g., Chap. 5 of Ref. \cite{Shk84} and references
therein): let's consider a Gaussian white noise $\eta(x)$ extending over all
the real axis, with zero mean and second moment $\langle \eta(x)\eta(y)
\rangle =\delta(x-y)$. Then for a given function $w(x)$ one defines
\begin{equation}\label{e2d}
U_g (x)=\frac{\hbar^2\sqrt{\sigma}}{m}\int_0^Lw(x-y)\,\eta(y) \, {\rm d}y \; ,
\end{equation}
where $\sigma$ is a parameter characterizing the disorder and whose meaning is
explained below.  If $w$ were a delta function, then $U_g$ would be a Gaussian
white noise over $[0,L]$ (and zero everywhere else). The actual function
$w(x)$ has a finite extension, and this induces finite correlations in the
disordered potential.

From (\ref{e2d}) it is clear that $\langle U_g \rangle =0$. If the domain of
integration in the r.h.s. of (\ref{e2d}) were extended to all $\mathbb{R}$,
$U_g$ would have a Gaussian distribution
\begin{equation}
{\cal P}(U_g)=\frac{\exp[-\frac{U_g^2}{2\,\Sigma^2}]}{\sqrt{2\pi\Sigma^2}} \; ,
\end{equation}
where
\begin{equation}\label{s2}
\Sigma^2=\sigma\, (\hbar^2/m)^2
\int_{\mathbb{R}}\!w^2(x){\rm d}x
\; .
\end{equation}
Defining the correlation function $C$ as
\begin{equation}\label{e2f0}
\langle U_g(x)U_g(x')\rangle - \langle U_g \rangle^2 = 
\sigma\,(\hbar^2/m)^2
\, C(x-x') \; ,
\end{equation}
one would get in this case
\begin{equation}\label{e2f}
C(x)=\int_{\mathbb{R}}\!\! {\rm d}y \, w(x+y) \, w(y) \; ,
\end{equation}
with a Fourier transform
\begin{equation}\label{e2g}
\hat{C}(q)=\int_{\mathbb{R}}\!\! {\rm d}x \, C(x) \,
\ep^{- {\rm i}\,q\,x}=\left|\hat{w}(q)\right|^2 \; ,
\end{equation}
where $\hat{w}$ is the Fourier transform of $w$. 

Imposing here the normalization condition
\begin{equation}\label{e2cbis}
\int_{\mathbb{R}}w(x)\, {\rm d}x =1 \; ,
\end{equation}
leads to a two-point correlation function (\ref{e2f0}) which is -- as in
Eq.~(\ref{e2b}) -- of the form of $\sigma(\hbar^2/m)^2$ multiplied by a
function whose integral over $x$ equals unity [$C(x)$ in (\ref{e2f0}) instead
  of $\delta(x)$ in (\ref{e2b})]. Thus, with definition (\ref{e2d}) and
normalization (\ref{e2cbis}), $\sigma$ plays for disorder (\ref{e2d})
the same role as $n_\delta/b^2$ [Eq.~(\ref{e2c})] for disorder
(\ref{e2a}): it characterizes the amplitude of the fluctuations of the
potential. The typical extent of $w(x)$ will in turn characterize the range of
the correlations.

Since $U_g$ as given by Eq.~(\ref{e2d}) is typically non zero only over a
region of finite extent, Eqs.~(\ref{e2f0}) and (\ref{e2f}) are only correct if
$x$ and $x'$ are inside this region. More precisely, they should be in this
region, at a distance from 0 or $L$ larger than the typical extent $\ell_c$ of
the function $w$. In the following we always consider the case where $L$ is
very large compared to $\ell_c$ (otherwise one could simply not speak of a
disordered region) and it is clear that the characteristics of the disorder
are properly defined only inside the disordered region.

We consider two special cases of correlation corresponding to different forms
of $w$: a Lorentzian
\begin{equation}\label{e2i}
w_{\sss L}(x)=\frac{1}{\pi} \,
\frac{\ell_c/2}{(\ell_c/2)^2+x^2} \; ,
\end{equation}
and a Gaussian
\begin{equation}\label{e2h}
w_{\sss G}(x)= \frac{1}{\ell_c\sqrt{\pi}}
\exp\left(-\frac{x^2}{\ell_c^2}\right) \; .
\end{equation}
We denote the corresponding potentials by $U_{\sss L}$ and $U_{\sss G}$. 
For the correlation functions one gets respectively
\begin{equation}\label{e2j}
C_{\sss L}(x)=
\frac{\ell_c/\pi}{\ell_c^2+x^2}\; , \quad
\hat{C}_{\sss L}(q)=
\ep^{-\ell_c|q|}\; ,
\end{equation}
and
\begin{equation}\label{e2k}
C_{\sss G}(x)=
\frac{\ep^{-x^2/(2\,\ell_c^2)}}
{\sqrt{2\,\pi\,\ell_c^2}} \; , \quad
\hat{C}_{\sss G}(q)=
\ep^{-q^2\ell_c^2/2} \; .
\end{equation}
In both cases  $\ell_c$ is the typical correlation radius.

The choice of a Lorentzian correlated disordered potential originates from
experimental and theoretical results in the case of micro-fabricated
guides. In this type of setting, the atoms are magnetically guided over a chip
\cite{fragmen}.  Unavoidable imperfections and irregularities in the design of
the circuit induce fluctuations in the current which, in turn, result in a
random contribution to the magnetic field used for guiding the atoms. Thus the
potential seen by the atoms has a random component which is typically
Lorentzian correlated, with a correlation length $\ell_c$ which decreases when
the distance between the guide and the chip increases
\cite{Kraft,Aspect_corrug,Wang2004,Pau05}. The Gaussian correlated potential
$U_{\sss G}$ is more academic but, by comparison with the results obtained
with $U_{\sss L}$ it allows one to check what is really specific to the
Lorentzian case, and what is a mere effect of finite correlation length.

\subsection{Speckle potential}\label{speckle}

Another experimentally relevant type of disorder is the so called speckle
potential which is generated by an optical speckle field produced by a laser
beam passing through a diffusing plate \cite{Goo07,Cle06,Fal08}.  The
corresponding potential will be denoted by $U_{\sss S}$ and may be
mathematically generated as follows \cite{rem_speckle}:
\begin{equation}\label{e2l}
U_{\sss S}(x)= \frac{\hbar^2 \sqrt{\sigma}}{m}
\left| \int_0^L\!\!\!\! w_{\sss S}(x-y) 
\left[\eta_1(y)+{\rm i}\,\eta_2(y)\right]  
{\rm d}y \right|^2 ,
\end{equation}
where $\eta_1$ and $\eta_2$ are two independent Gaussian white noise processes
of zero mean with $\langle\eta_\alpha(x) \eta_{\alpha'}(x')\rangle =
\delta(x-x') \delta_{\alpha \alpha'}$
($\alpha$ and $\alpha'=1$ or 2).

Here also we characterize the disorder by studying its statistical
properties
in the limit where the domain of integration in the r.h.s. 
of (\ref{e2l}) is extended to all $\mathbb{R}$. In this case one gets
\begin{equation}
{\cal P}(U_{\sss S})=\frac{\exp(-\frac{U_{\sss S}}{2\Sigma^2})}{2\Sigma^2}
\; ,
\end{equation}
where $\Sigma$ is given by formula (\ref{s2}) (replacing $w$ by $w_{\sss S}$).
This yields $\langle U_{\sss S} \rangle=2\Sigma^2$ and the correlation
function defined in Eq.~(\ref{e2f0}) reads here
\begin{equation}\label{e2m}
\begin{split}
C_{\sss S}(x-x')= & 
\frac{1}{\sigma\,(\hbar^2/m)^2}
\left[
\langle U_{\sss S}(x)U_{\sss S}(x')\rangle - 
\langle U_{\sss S} \rangle^2 \right] \\
= & 4 \left[
\int_{\mathbb{R}}\!\!{\rm d}y\, w_{\sss S}(x-x'+y)w_{\sss S}(y)
\right]^2  \; .
\end{split}
\end{equation}
Contrarily to the choice (\ref{e2cbis}), $w_{\sss S}$ should not be
normalized to unity here because -- from Eq.~(\ref{e2l}) -- this is
homogeneity-wise impossible. Instead, the choice
\begin{equation}\label{ws}
w_{\sss S}(x)=
\left(\frac{\ell_c}{4\,\pi^3}\right)^{1/4} \,
\frac{\sin(\frac{x}{\ell_c})}{x} \; ,
\end{equation}
corresponds to the typical experimental situations 
\cite{Cle06} and leads to a correlation function
\begin{equation}
C_{\sss S}(x)=\frac{\ell_c}{\pi}
\frac{\sin^2(\frac{x}{\ell_c})}{x^2} \; ,
\end{equation}
whose integral over $x$ equals
unity and whose Fourier transform is
\begin{equation}\label{csq}
\hat{C}_{\sss S}(q)=\left\{
\begin{array}{cl}
1-|q|\,\ell_c/2 
& \mbox{if}\;\;\; |q|\,\ell_c<2\; , \\
& \\
0 & \mbox{otherwise}\; . \end{array}
\right.
\end{equation}
Hence, definition (\ref{e2l}) and choice (\ref{ws}) correspond here
also to characterizing the amplitude of the disorder's fluctuations by the
parameter $\sigma$ and the range of the correlations by $\ell_c$.

\section{Supersonic stationary regime}\label{ssr}

As explained in Ref.~\cite{Pau07}, and recalled in the Introduction, Anderson
localization in a weakly repulsive Bose-Einstein condensate is only possible
in the supersonic regime (cf. Fig. \ref{fig1}) which we consider now. In the
present section we first analyze the transmission across a short disordered
sample, in which case perturbation theory is applicable (Sec. \ref{pts}). We
then turn to generic non-perturbative configurations (Sec. \ref{non-pert})
where Anderson localization is expected. In this regime we obtain evidences of
the occurrence of Anderson localization in the presence of interaction.
Finally we discuss the upper limit of the localized regime and the onset of
time dependent flows for long disordered samples (Sec. \ref{sec-lstar}).

\subsection{Perturbation theory ($\lambda\ll 1$)}\label{pts}

In the supersonic stationary regime, simple perturbation theory yields
$n(x,t)=n_0+\delta n(X)$ where \cite{Leb01}
\begin{equation}\label{spt1}
\delta n(X)=\frac{2\,m\,n_0}{\hbar^2\,\kappa}
\int_{-\infty}^X\!\!\!\!{\rm d}y \, U(y)\,
 \sin[2\kappa(X-y)] \; ,
\end{equation}
and $\kappa$ is given by Eq.~(\ref{pt4}). Perturbation theory always predicts
a stationary density profile. This is certainly wrong when $V$ is close to $c$
(cf. Fig. \ref{fig1}), but in this case $\kappa$ gets very small and one
precisely goes out of the domain of validity of perturbation theory ($\delta
n$ as given by (\ref{spt1}) is no longer small compared to $n_0$). 

Far ahead of the obstacle (in a region where $X-L$ is larger than $\ell_c$ and
$\kappa^{-1}$) (\ref{spt1}) gives
\begin{equation}\label{spt2}
\frac{\delta n(X)}{n_0}=\frac{2 m}{\hbar^2 \kappa} \,\mbox{Im}\, \left\{
\ep^{2 {\rm i} \kappa X} \hat{U}(2 \kappa ) \right\} \; .
\end{equation}
The perturbative regime in which Eqs. (\ref{spt1},\ref{spt2}) are valid is
also the one where the constant $\lambda$ in (\ref{t5},\ref{t6}) is small
compared to unity. This can be inferred from the comparison of (\ref{spt2})
and (\ref{t6}) which indeed shows that $\lambda\ll 1$ in the regime where
(\ref{spt2}) holds and that, in this case, $\sqrt{\lambda}\simeq m
|\hat{U}(2\kappa)|/(\hbar^2\kappa)$.  The corres\-pon\-ding reflection
coefficient can then be obtained from (\ref{t7}), yielding
\begin{equation}\label{spt3}
R\simeq \lambda \simeq 
\frac{m^2}{\hbar^4 \kappa^2} \, |\hat{U}(2\kappa)|^2
\; .
\end{equation}
From (\ref{spt3}) it is clear that the average reflection coefficient is
\begin{equation}\label{spt5}
\langle R\, \rangle =\langle \lambda\, \rangle =
\frac{m^2}{\hbar^4\kappa^2} \langle |\hat{U}(2\kappa)|^2\rangle 
\ll 1 \; .
\end{equation}
Furthermore, one can show that the corresponding probability distribution of
the reflection coefficient is Poissonian with
\begin{equation}\label{spt4}
P(R)=\frac{1}{\langle R\, \rangle } \exp (-R/\langle R\, \rangle) \; .
\end{equation}
Note that for properly normalizing this probability distribution for
$R\in[0,1]$, one should include in the prefactor of the r.h.s. of (\ref{spt4})
a correcting term of order $\exp(-1/\langle R\, \rangle)$ which can be safely
neglected in the limit (\ref{spt5}).

\begin{figure}
\begin{center}
\includegraphics[width=0.95\columnwidth]{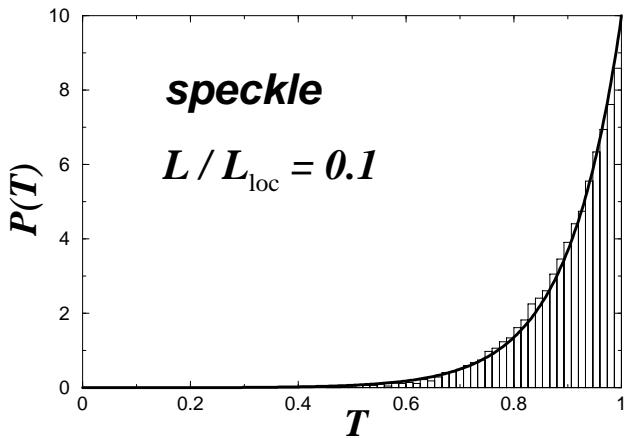}
\end{center}
\caption{Probability distribution $P(T)$ for the transmission coefficient $T$
  in a potential $U_{\sss S}$ with $\sigma=3.14\,\mu^2\xi$, $\ell_c=0.1\xi$
  and $L=50\,\xi$ moving at velocity $V=7c$ in a condensate of initial
  constant density $n_0\,\xi=1$. The corresponding localization length is
  $L_{\rm loc}=500\,\xi$. The histogram corresponds to a statistical analysis
  of the results of the numerical solution of Eq.~(\ref{e1}) for 10000
  different random potentials. The solid line is the perturbative result
  (\ref{e7b}).\label{fig4}}
\end{figure}

We give in Appendix \ref{app1} a demonstration of result (\ref{spt4}) for
the special case of a correlated Gaussian potential $U_g$ of type
(\ref{e2d}). Below, we show that the same result holds for a potential
$U_\delta$ of type (\ref{e2a}) [see Eq.~(\ref{dmpk0}) in section
  \ref{non-pert}], and we checked numerically that it is also the case for the
speckle potential $U_{\sss S}$ (\ref{e2l}) (cf Fig.~\ref{fig4}). In all these
cases, the average reflection coefficient reads (up to the above discussed
exponentially small correction)
\begin{equation}\label{spt10}
\langle R \, \rangle =L/L_{\rm loc}(\kappa) \; ,
\end{equation}
where 
\begin{equation}\label{spt11}
L_{\rm loc}(\kappa)
=\frac{\kappa^2/\sigma}{\hat{C}(2\kappa)}\; .
\end{equation}
We recall that the function $\hat{C}$ depends on the type of disorder
considered. For a potential of form (\ref{e2a}) one has
$\hat{C}_\delta\equiv 1$, for the other potentials considered in this work it
is given by (\ref{e2j}), (\ref{e2k}) and (\ref{csq}).

Concomitantly to distribution (\ref{spt4}) of reflection coefficients one
gets for the transmission
\begin{equation}\label{e7b}
P(T)=\frac{L_{\rm loc}}{L} \exp\left\{
-(1-T) \frac{L_{\rm loc}}{L} \right\}\; .
\end{equation}
From (\ref{e7b}) [or (\ref{spt10})] one gets
\begin{equation}\label{e7}
\langle T\rangle = 1 - \frac{L}{L_{\rm loc}(\kappa)} \; .
\end{equation}
The perturbative approach holds when $\langle R\,\rangle\ll 1$ i.e., when
$L\ll L_{\rm loc}$. This corresponds to the region which is denoted as
``ballistic'' in Fig. \ref{fig1} \cite{ohm}. Its accuracy is shown for
$L/L_{\rm loc}=0.1$ in Fig.~\ref{fig4} for a speckle potential $U_{\sss S}$ of
type (\ref{e2l}) (we also checked this prediction for the potentials
$U_{\delta}$ and $U_{\sss G}$, with also excellent results).

At this stage, $L_{\rm loc}$ is simply a notation for expression
(\ref{spt11}), but it will be shown to be the actual localization length of
the matter wave in a disordered potential (in section \ref{non-pert}).  

The results derived here also hold for a noninteracting gas, obtained by
taking the limit $g\rightarrow 0$ in Eq.~(\ref{e1}), in which case $c=0$ and
$\kappa=k$.  Eq.~(\ref{spt11}), with $\kappa$ replaced by $k$, then coincides
with the Antsygina-Pastur-Slyusarev formula for the localization length
\cite{Ant81,Lif88} and the distribution of transmissions (\ref{e7b}) holds,
with $L_{\rm loc}=L_{\rm loc}(k)$.

\begin{figure}
\begin{center}
\includegraphics*[width=0.95\columnwidth]{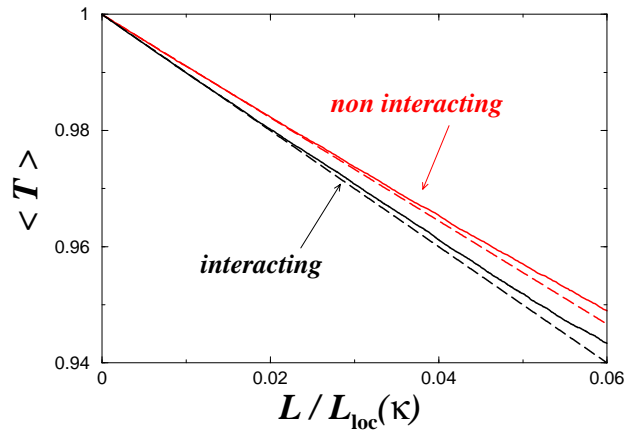}
\end{center}
\caption{(Color online): Average transmission as a function of $L$ for a
  potential $U_\delta$ (characterized by $n_\delta \xi=0.5$ and
  $\xi/b=0.1$). In the interacting case $V=3\,c$ and $L_{\rm loc}(\kappa)=1600
  \, \xi$.  The non-interacting case is drawn for the same velocity and
  corresponds to a value $L_{\rm loc}(k)=\frac{9}{8} L_{\rm
    loc}(\kappa)=1800\,\xi$. In both cases the dashed line is the analytical
  result (\ref{e7}) and the solid line corresponds to a statistical analysis
  of the results of the numerical solution of Eq.~(\ref{e1}) for 15000
  different random potentials. The departure of the numerical results from the
  dashed lines occurs when the systems leaves the perturbative
  regime.\label{fig5}}
\end{figure}

In the present work, those formulas are modified to include interactions.  The
generalization simply consists in replacing the wave vector $k=mV/\hbar$ by
$\kappa =m(V^2-c^2)^{1/2}/\hbar$.  This replacement has, as an important
physical consequence, the effect of diminishing, at a given speed $V$, the
localization length (there is an effective reduction of the available kinetic
energy by the repulsive interactions). For instance, in the case of a
potential $U_\delta$, since $L_{\rm loc}(\kappa)\propto \kappa^2$, there is a
relative difference $c^2/V^2$ between $L_{\rm loc}(\kappa)$ and $L_{\rm
  loc}(k)$, that is 11 $\%$ for $V=3\, c$. This is illustrated in
Fig.~\ref{fig5}, which displays the average $\langle T\,\rangle$ as a function
of $L$ for a disorder $U_\delta$ of type (\ref{e2a}), with and without
interactions.

\subsection{Non perturbative approach}\label{non-pert}

When the size $L$ of the sample is large compared to the value $L_{\rm loc}$
determined in Section \ref{pts}, the perturbative approach fails. We now
propose a non perturbative method allowing to treat both the regimes $L<L_{\rm
  loc}$ and $L\ge L_{\rm loc}$ and showing that $L_{\rm loc}$, as defined in
Eq.~(\ref{spt11}), is indeed the localization length in the presence of
interactions.

Within the framework of the non perturbative approach, we are able to
provide approximate analytical results in the case of the model disorder
potential (\ref{e2a}). This potential being zero between two impurities, one
can write a series of first integrals of Eq.~(\ref{e4}) in each segment
$]x_n,x_{n+1}[$ as follows~:
\begin{equation}\label{np1}
\frac{\hbar^2}{2 m}\left(\frac{{\rm d}A}{{\rm d}X}\right)^2 + W[A(X)]
=E^{(n)}_{\rm cl} \; .
\end{equation}
In the region $X<x_1=0$ the integration constant $E_{\rm cl}^-$ of
Eq.~(\ref{t1}) is denoted as $E_{\rm cl}^{(0)}$ in (\ref{np1}) (taking
$x_0=-\infty$) and is zero, whereas in the region $X>x_{N}$ one has
$E_{\rm cl}^+= E_{\rm cl}^{(N)}$ (and $x_{N+1}=+\infty$).

\begin{figure}
\begin{center}
\includegraphics[width=0.95\columnwidth]{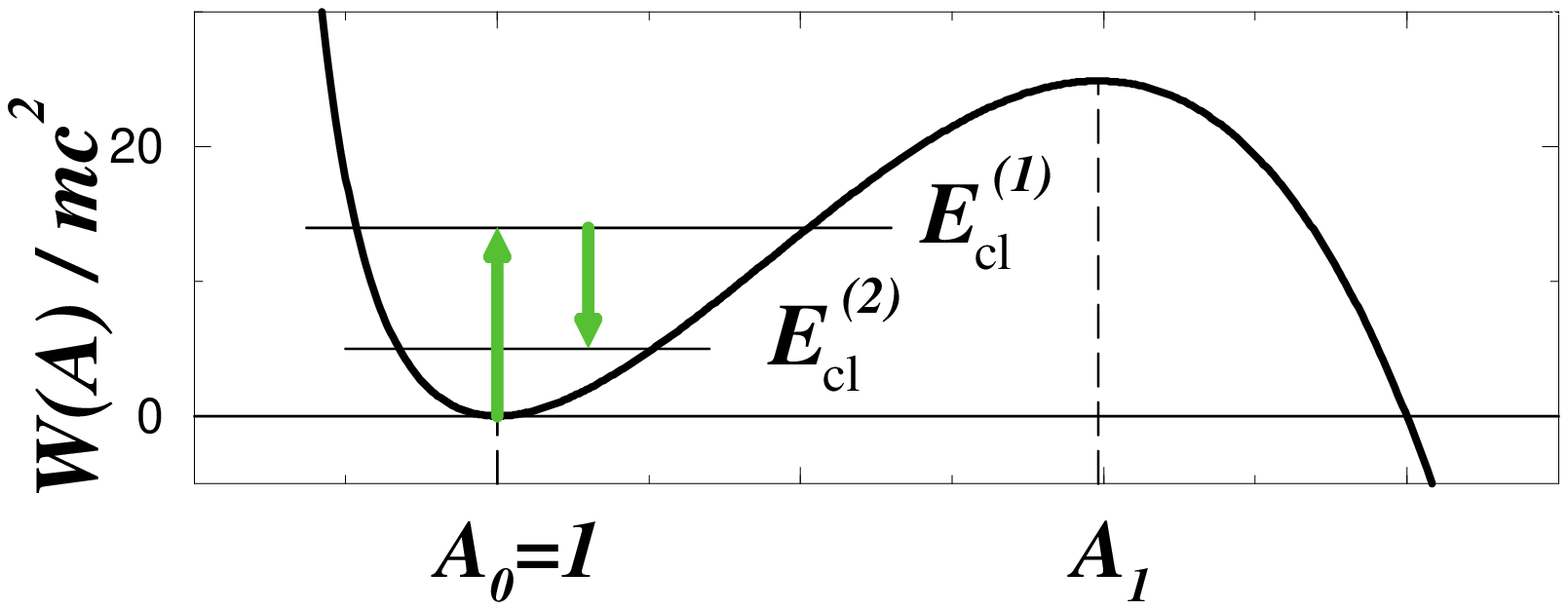}
\includegraphics[width=0.95\columnwidth]{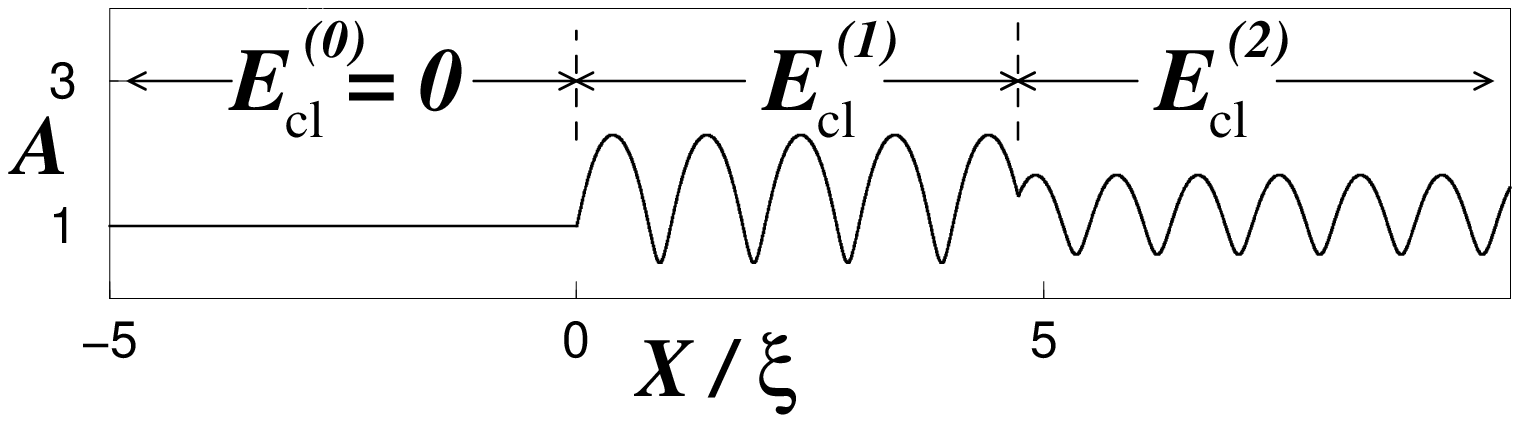}
\end{center}
\caption{(Color online) Upper panel: $W$ as a function of $A$ (drawn for
  $V/c=4$). For $X<x_1=0$, the fictitious particle is initially at rest at the
  bottom of potential $W$ with $E_{\rm cl}^{(0)}=0$. The value of the
  classical energy changes from $E_{\rm cl}^{(n-1)}$ to $E_{\rm cl}^{(n)}$ at
  each impurity $x_n$. The lower panel displays the corresponding oscillations
  of $A(X)$, with two impurities at $x_1 = 0$ and $x_2 = 4.7 \, \xi$ (their
  position is indicated by vertical dashed lines).\label{fig6}}
\end{figure}

From (\ref{e4}) it is a simple matter to show that the matching condition of
the density at impurity position $x_n$ is 
\begin{equation}\label{np1a}
A'(x_n^+)-A'(x_n^-)=\frac{2}{b}\, A(x_n) \; ,
\end{equation}
where $A'(x_n^-)$ [$A'(x_n^+)$] denotes the limit of the derivative ${\rm
  d}A/{\rm d}X$ at the left [at the right] of $x_n$. Relation
(\ref{np1a}) between the derivatives of the amplitude results [from Eq.
  (\ref{np1})] in a relation between the classical energies:
\begin{equation}\label{np2}
E_{\rm cl}^{(n)}=E_{\rm cl}^{(n-1)}+\frac{2 \hbar^2}{m b^2} \, A(x_n)\left[
b \, A'(x_n^-) + A(x_n)\right] \; .
\end{equation}

Hence, Eq.~(\ref{np1}) allows to draw a classical analogous of the solution of
the nonlinear Sch\"odinger equation in the presence of potential (\ref{e2a})
formed by a series of delta peaks: the fictitious classical particle defined
in Sec. \ref{transmis} evolves in the potential $W$ and experiences kicks at
``times'' $x_n$. Each kick changes the ``energy'' according to (\ref{np2}), as
illustrated in Fig. \ref{fig6}. The key point in the remaining of this section
will be to derive the probability distribution of $E_{\rm cl}^{+}=E_{\rm
  cl}^{(N)}$ which then directly allows one to get the distribution of
$\lambda$'s and of the transmission coefficients [through Eqs. (\ref{t5}) and
  (\ref{t7})].

Let us introduce the quantities
\begin{equation}\label{np2a}
\lambda_n=\frac{m E_{\rm cl}^{(n)}}{2\, \hbar^2 \kappa^2}
\; , \quad\mbox{and}\quad
\Lambda_n=\sqrt{(\lambda_n)^2+\lambda_n} \; .
\end{equation}
The parameters $\lambda$ and $\Lambda$ defined in (\ref{t5}) and (\ref{t6b})
are related to the ones of Eq.~(\ref{np2a}) by $\lambda=\lambda_{N}$ and
$\Lambda=\Lambda_{N}$ (i.e., $\lambda$ is the last of $\lambda_n$'s; the same
holds for $\Lambda$). Denoting by $\theta_{n-1}$ the value of the phase
$\theta$ [appearing in(\ref{t6})] for $X\in]x_{n-1},x_{n}[$, one gets in this
    domain [the derivation is exactly the same as for Eq.~(\ref{t6})]
\begin{equation}\label{t6bis}
A^2(X)=1+2 \lambda_{n-1}+2 \Lambda_{n-1}
\cos\left(2\kappa X+\theta_{n-1}\right) \; ,
\end{equation}
and one can rewrite Eq.~(\ref{np2}) as
\begin{eqnarray}\label{np3}
E_{\rm cl}^{(n)} & = & E_{\rm cl}^{(n-1)} +   
\frac{2\hbar^2}{m b^2} \left[1+2\,\lambda_{n-1}\right] \nonumber \\ 
& + & \frac{4\hbar^2}{m b^2} \Lambda_{n-1} \sqrt{\kappa^2 b^2 +1}  
\; \zeta_{n-1}
\; ,
\end{eqnarray}
where
\begin{equation}\label{np4b}
\zeta_{n-1}=\cos\left[ 2\,\kappa\, x_n + \theta_{n-1} 
+ \tan^{-1}\left(\kappa b \right)\right] \; .
\end{equation}
Using definition (\ref{np2a}) one can rewrite
Eq.~(\ref{np3}) in terms of the parameter $\lambda_n$ as
\begin{equation}\label{np5}
\lambda_n=\lambda_{n-1}+\frac{1+2\,\lambda_{n-1}}{\kappa^2 b^2} + 
\frac{2\,\Lambda_{n-1}}{\kappa^2 b^2} \sqrt{\kappa^2 b^2 +1} \; 
\zeta_{n-1} \; .
\end{equation}

Eqs.~(\ref{np3},\ref{np5}) are valid provided (\ref{t2}) holds, {\it
  i.e.}, provided $E_{\rm cl}^{(n-1)}\ll W(A_1)$, which reads
\begin{equation}\label{np5b}
E_{\rm cl}^{(n-1)}\ll \frac{\hbar^2\kappa^2}{m} \,(\kappa^2\,\xi^2) \; ,
\quad \mbox{{\it i.e.}}\; , \quad 
\lambda_{n-1} \ll \kappa^2\,\xi^2 \; .
\end{equation}
In the following we also impose the condition
\begin{equation}\label{np4a}
\kappa \, b \gg 1 \; .
\end{equation}
Precisely, we neglect all the quantities of order $1/(\kappa^3 b^3)$.  This is
an important technical point. It facilitates the analysis by allowing one to
get simple formulas as we now illustrate in the perturbative case:
Eq.~(\ref{np5}) allows for instance to compute the average value of the
reflection coefficient in the perturbative regime (as already done in Section
\ref{pts}, Eq.~(\ref{spt10})). In this regime, additionally to condition
(\ref{np5b}) one has $\lambda_{n}\ll 1$. Then Eq.~(\ref{np5}) implies at
leading order $\langle \lambda_{n} \rangle= \langle\lambda_{n-1}\rangle +
1/(\kappa^2 b^2)$ which, together with the initial condition $\lambda_0=0$,
leads immediately to
\begin{equation}\label{npp}
\langle \lambda \rangle = \frac{1}{\kappa^2 b^2}\, \langle N \rangle 
= \frac{1}{\kappa^2 b^2}\,\frac{ L}{l_\delta} \; ,
\end{equation}
which is identical to result (\ref{spt10}) in the case of a potential
$U_\delta$ for which $L_{\rm loc}(\kappa) = \kappa^2/\sigma =
\kappa^2b^2\, l_\delta$ [cf. Eq. (\ref{spt11})].

Let us now proceed and consider the generic non perturbative regime where
$\lambda_{n}$ may become large compared to unity and where (\ref{t2}) and
(\ref{np5b}) are still valid. Taking the condition (\ref{np4a})
into account, Eq.~(\ref{np5}) reads
\begin{equation}\label{np6}
\lambda_n=\lambda_{n-1}+\frac{1+2\,\lambda_{n-1}}{\kappa^2 b^2} + 
\frac{2\,\Lambda_{n-1}}{\kappa\, b} \; 
\zeta_{n-1} \; .
\end{equation}
It is natural to assume that the phase of the cosine in the r.h.s. of
(\ref{np4b}) is uniformly distributed in $[-\pi,\pi]$ and independent of the
phase at step $n-1$.  This could be called a ``phase randomization''
approximation. This relies on hypothesis
(\ref{np4a}) and on the assumption that there is
a large number of density oscillations over the (random) length between
$x_{n-1}$ and $x_{n}$, i.e.,
\begin{equation}\label{gnp2bis}
\kappa\langle x_n-x_{n-1}\rangle=\kappa\,l_\delta \gg 1 \; .
\end{equation}
Then, the argument of the cosine in definition (\ref{np4b}) is uniformly
distributed, $\zeta_n$'s are uncorrelated random variables, with all the
same law characterized by its average $\langle \zeta_n\rangle = 0$ and variance
\begin{equation}\label{varxi}
\langle
\zeta_n \zeta_{n'}\rangle =\frac{1}{2}\delta_{n,n'} \; .
\end{equation}
Note that the regimes (\ref{np4a}) 
and (\ref{gnp2bis}) imply that
\begin{equation}\label{npzut}
\frac{\hbar^2\kappa^2}{2\, m}\gg \frac{\hbar^2\,n_\delta}{m\,b}=
\langle U_\delta\rangle \; ,
\end{equation}
which in turn implies that the kinetic energy $\frac{1}{2}m V^2$ is much
larger than $\langle U_\delta\rangle$; {\it i.e.}, one is exactly in the
Anderson regime where the incident kinetic energy is much larger than the
typical value of the (disordered) potential representing the obstacle. Hence a
classical particle would flow almost unperturbed
over the potential but, as we shall see, a
quantum particle experiences an exponentially small transmission.

Let $P(\lambda,n)d\lambda$ be the probability that $\lambda_n$ lies in the
interval $\lambda$, $\lambda+{\rm d}\lambda$. Going to the continuous limit
and defining the continuous variable $t=n/(\kappa^2 b^2) = X/L_{\rm
  loc}$ [where $L_{\rm loc}=\kappa^2/\sigma$ is the parameter (\ref{spt11}) in
  the case of a potential $U_\delta$] it is shown in Appendix \ref{app2} that
$P(\lambda,t)$ verifies the following
Fokker-Planck equation
\begin{equation}\label{dmpk}
\frac{\partial P}{\partial t}=
\frac{\partial }{\partial \lambda}\left[\lambda(1+\lambda)
\frac{\partial P}{\partial \lambda}\right]\; .
\end{equation}
Equation (\ref{dmpk}) follows directly from Eq.~(\ref{np6}) in the regime
where conditions (\ref{np4a}) and (\ref{gnp2bis}) hold. It is precisely the
Dorokhov-Mello-Pereyra-Kumar (DMPK) equation \cite{papierdmpk} for the
transmission in a single disordered channel [with $T=1/(1+\lambda)$].
Equation (\ref{dmpk}) is sometimes referred to as Mel'nikov's equation (after
Ref. \cite{Mel80}) but has a much longer history (see the discussion in
Refs. \cite{Lif88,Bee97}).

Since before entering the disordered region the particle has a classical
energy $E_{\rm cl}^{(0)}=0$ corresponding to $\lambda=0$, Eq.~(\ref{dmpk}) has
to be solved for the initial condition
\begin{equation}\label{np8}
\lim_{t\to 0}P(\lambda,t)=\delta_{+}(\lambda) \; ,
\end{equation}
where $\delta_{+}$ is the one-sided delta function:
$\int_0^\infty\delta_{+}(\lambda){\rm d}\lambda=1$. In the limit of small $t$
(i.e., in the perturbative regime $X\ll L_{\rm loc}$), 
$\lambda$ remains small and one can approximate in the
r.h.s. of (\ref{dmpk}) the term $\lambda(\lambda+1)$ by $\lambda$. It is then
simple to verify that the solution of this approximate equation that satisfies
(\ref{np8}) is
\begin{equation}\label{dmpk0} 
P(\lambda,t) = \frac{\exp\{-\lambda/t\}}{t} 
\quad \mbox{for} \quad
t\ll 1 \; .
\end{equation}
This result for the small $t$ solution of the DMPK equation has been already
obtained in Ref.~\cite{Mel86} (see also the discussion in
Ref.~\cite{Bee94}). The distribution law (\ref{dmpk0}) is exactly equivalent
to distribution (\ref{spt4}) of the reflection coefficient in the
perturbative regime and this proves the validity of the Poissonian
distribution (\ref{e7b}) for a potential $U_\delta$ of type (\ref{e2a}) 
\cite{rem7}.

In the general case (i.e., for all $t\ge 0$) the solution of (\ref{dmpk}) with
the initial condition (\ref{np8}) is (see, {\it e.g.},
Refs.~\cite{Lif88,Bee97} and references therein)
\begin{equation}\label{np9}
P(\lambda,t)=\frac{\ep^{-t/4}}{\sqrt{2\,\pi\,t^3}}
\int_{u_\lambda}^{\infty}  
\frac{u\;\ep^{-u^2/(4\,t)}}{\sqrt{\cosh(u)-1-2\,\lambda}} \; {\rm d}u
\; ,
\end{equation}
where $u_\lambda=\cosh^{-1}(1+2\lambda)$.

From distribution (\ref{np9}), a lengthy computation or alternatively the
direct use of the DMPK equation (\ref{dmpk}) \cite{Abr81} yields
\begin{equation}\label{np10}
\langle \, \ln T \rangle =
\int_0^{\infty}\!\!{\rm d}\lambda \, \ln\left(\frac{1}{1+\lambda}\right)
P(\lambda,t)= -t \; .
\end{equation}

In the large $t$ limit, distribution (\ref{np9}) tends to a log-normal
distribution, i.e., the distribution of the variable $\ln T$ is Gaussian
(see Ref.~\cite{Abr81})
\begin{equation}\label{np11}
P(\ln T,t)=
\frac{\exp\left\{-(t+\ln T)^2/4 t\right\}}{\sqrt{4\pi t}} 
\quad \mbox{for} \quad
t\gg 1 \; .
\end{equation}
From this distribution one gets the correct average $\langle \ln T\rangle =
-t$, Eq.~(\ref{np10}), and a standard deviation $[ \langle (\ln T)^2\rangle -
  \langle \ln T\rangle^2]^{1/2} = \sqrt{2 t}$, which is in agreement with the
exact result in the limit $t\gg 1$ \cite{Abr81}.  At the extremity of a sample
of length $L$ one has $t=L/L_{\rm loc}$, and the distribution (\ref{np11}) is
the log-normal distribution of transmission typical for Anderson localization
in the regime $L\gg L_{\rm loc}$ (see, e.g., Ref.~\cite{Tig99,Bee97}). As a side
product of this analysis, Eqs.~(\ref{np10}) and (\ref{np11}) confirm that
$L_{\rm loc}$ is indeed the localization length as was anticipated in the
notation.

\begin{figure}
\begin{center}
\includegraphics[width=0.95\columnwidth]{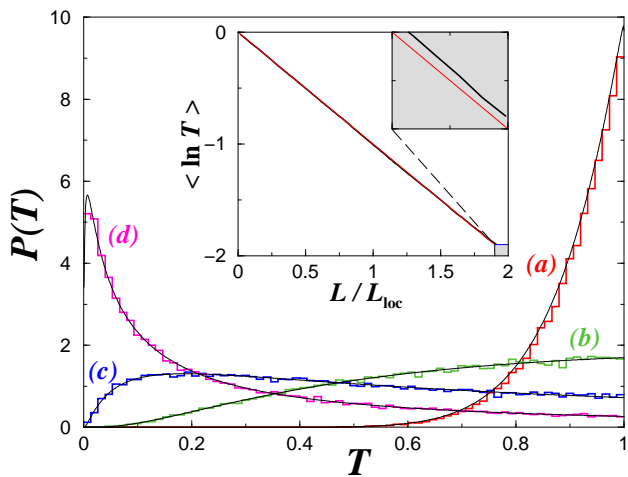}
\end{center}
\caption{(Color online) Probability distribution of the transmission through a
  disordered potential $U_\delta$ of type (\ref{e2a}) (characterized by
  $\xi/b=0.5$ and $n_\delta\xi=0.5$) plotted for different values of the ratio
  $t=L/L_{\rm loc}(\kappa)$ with $V=30 \,c$. The black solid lines are the
  DMPK result (\ref{np9}) and the colored histograms correspond to the
  numerical simulations (50000 samples used for each value of $t$).  Cases
  (a), (b), (c) and (d) correspond respectively to $t=0.1$, 0.5, 1 and 2. The
  inset displays $\langle \ln T\rangle$ as a function of $t$. The thick solid
  line is extracted from numerical simulations and the thin (red) solid line
  is the DMPK prediction (\ref{np10}). They can be distinguished only around
  $t\simeq 2$ as shown in the blowup of the (gray) shaded region for
  $1.9<t<2$.
\label{fig7}}
\end{figure}

We have tested
the validity of the DMPK approach for a Bose-Einstein beam of
interacting particles propagating in a disordered potential $U_\delta$ of type
(\ref{e2a}). The numerical results for the probability distribution $P(T)$ are
compared on Fig. \ref{fig7} with the DMPK prediction (\ref{np9}). The
agreement is seen to be excellent. The distribution evolves from the
Poissonian result (\ref{e7b}) (for low values of $L/L_{\rm loc}$) towards a
distribution peaked at low $T$-values for large $L/L_{\rm loc}$. In this
latter case one can check that the distribution tends to a log-normal by
plotting $P(\ln T)$.

We have also checked the validity of the present approach over a sizable
range of lengths of disordered region and of intensities of disordered
potential by plotting in the inset of Fig. \ref{fig7} the average $\langle
\rm{ln}T\rangle$ as a function of $L/L_{\rm loc}$. The agreement of the
numerical results with the DMPK prediction (\ref{np10}) is excellent. Note
however the beginning of a small departure around $L/L_{\rm loc}\simeq 2$;
this effect will be studied more thoroughly in Sec. \ref{sec-lstar}
(cf. Fig. \ref{fig9}).

\begin{figure}
\begin{center}
\includegraphics[width=0.95\columnwidth]{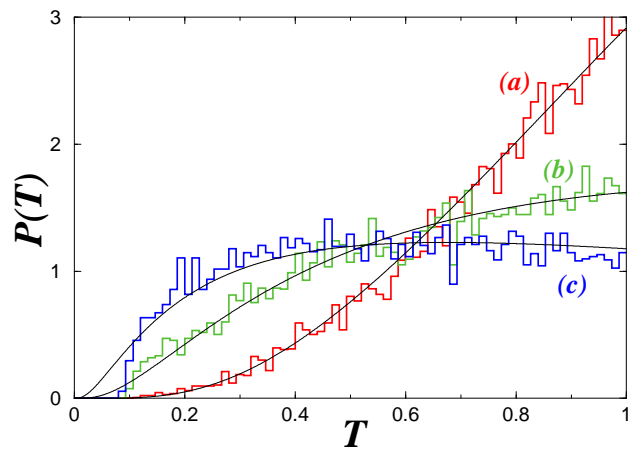}
\end{center}
\caption{(Color online) Probability distribution of the transmission through a
  speckle disordered potential $U_{\sss S}$ (characterized by
  $\ell_c/\xi=0.05$ and $\sigma=3.13$ $\mu^2\xi$) plotted for different values
  of the ratio $t=L/L_{\rm loc}(\kappa)$ with $V=13\, c$.  Curves (a), (b) and
  (c) corresponds respectively to $t=0.31$, 0.52 and 0.68.  For each curve,
  the black solid line is the DMPK result (\ref{np9}) and the colored
  histogram is the result of 10000 numerical simulations.\label{fig8}}
\end{figure}

Finally, we discuss numerical results obtained for the disordered potentials
introduced in Secs. \ref{gaussian} and \ref{speckle}.  Although we do not have
an analytical derivation of the DMPK equation for these potentials, the
numerical results indicate a very good quantitative agreement for a disordered
potential $U_{\sss G}$ and for a speckle potential $U_{\sss S}$. We display
the comparison of the numerical data with the DMPK predictions for a speckle
potential in Figure \ref{fig8}. The same agreement is obtained for a Gaussian
potential $U_{\sss G}$. Hence, the behavior analytically predicted for the
potential $U_\delta$ appears to be of general validity, meaning that the above
defined regime of ``phase randomization'' can probably be extended to
correlated potentials, leading to a regime of single parameter
scaling. However, we have noticed that, although showing an overall good
agreement with the DMPK prediction, the Lorentzian correlated potential
$U_{\sss L}$ exhibits some deviations in the tail of the distribution, the
details of which will be studied elsewhere.

\subsection{Threshold for the existence of a stationary flow}\label{sec-lstar}

In the previous sections the main effect of interaction has been shown
to be a renormalization of the localization length $L_{\rm loc}$. 
Interaction induces a modification of the wave vector: 
expression (\ref{spt11}) for the localization length coincides with the
noninteracting one but computed for an effective interaction-dependent wave
vector $\kappa$ given by Eq.~(\ref{pt4}), instead of $k=mV/\hbar$. The repulsive
interaction diminishes the available kinetic energy and therefore reduces the
localization length with respect to the noninteracting case (since
$\kappa<k$).

We now discuss another, more spectacular, effect of interactions on the
localization properties of a propagating BEC on a disordered potential.

In the previous sections \ref{pts} and \ref{non-pert}, we completely neglected
the presence of an upper limit for the classical energy $E_{\rm cl}$, which is
given by the local maximum of the effective potential $W(A)$, namely $E_{\rm
  cl}^{\rm max} = W(A_1)$ (see Fig. \ref{fig3}).  Trajectories that pass
beyond $E_{\rm cl}^{\rm max}$ would become unstable and develop singularities
with infinitely large density at $X\to\infty$.  In practice this implies, on
the level of Eq. (\ref{e1}), that a stationary flow cannot be maintained in
this case and that the disorder induces time-dependent dynamics of the
condensate.

In the vicinity of $E_{\rm cl}^{\rm max}$, the density profile of the
condensate in between two adjacent scatterers becomes quite different from the
cosine shape (\ref{t6}) that was derived for weak nonlinearities and/or low
density modulations, and resembles more to a periodic train of gray solitons
\cite{Leb01}. In a crude approximation, we neglect this complication and
assume that the spatial evolution of the density is still given by
Eq.~(\ref{t6}) for all classical energies until $E_{\rm cl} = E_{\rm cl}^{\rm
  max}$.  Trajectories that happen to pass beyond $E_{\rm cl}^{\rm max}$
are considered to be ``lost'', i.e., they do no longer contribute to
the probability distribution for the transmission.  This formally amounts to
introducing a ``sink'' in the stochastic equation (\ref{np6}), namely at
$\lambda = \lambda_{\rm max}=m\,E_{\rm cl}^{\rm max} / (2\hbar^2\kappa^2)$.
In the corresponding Fokker-Planck equation
(\ref{dmpk}), this sink is appropriately modeled by imposing the boundary
condition
\begin{equation} \label{bc}
	P(\lambda_{\rm max},t) = 0\; .
\end{equation}
As a consequence of this boundary condition, the integrated probability
distribution $\int_0^{\lambda_{\rm max}} P(\lambda,t) {\rm d} \lambda$ is no
longer conserved, but decreases with increasing $t$, i.e., increasing length
$L$ of the disorder region.

In the following, we show how this affects the DMPK predictions of section
\ref{non-pert} and how the ``survival probability'', i.e., the fraction of
trajectories that remain below this boundary at given length $L$, can be
analytically computed in the limit $V\gg c$. In this limit, from
Eq. (\ref{t1c}) one gets $E_{\rm cl}^{\rm max}\simeq mV^4/(8\,c^2)$ and thus
$\lambda_{\rm max}\simeq V^2/(16\,c^2)\gg 1$. Modifications of the probability
density $P(\lambda,t)$ due to the presence of the sink appear only when the
typical value of $\lambda$ is of order $\lambda_{\rm max}$ which, as just
remarked, is large compared to unity in the case $V\gg c$.  In this case
$P(\lambda,t)$ is already negligibly small around $\lambda \sim 1$.  We
therefore make the approximation $\lambda + 1 \simeq \lambda$ in the
Fokker-Planck equation (\ref{dmpk}), which then reads
\begin{equation}\label{dmpk-large}
\frac{\partial P}{\partial t}=
\frac{\partial }{\partial \lambda}\left[\lambda^2
\frac{\partial P}{\partial \lambda}\right]\; .
\end{equation}
Using, from now on, the probability distribution $P(\ln T, t)$ for finding a
given value of $\ln T$ at fixed $t \equiv L / L_{\rm loc}$, we obtain
in this limit
\begin{equation}\label{lognorm-eq}
\frac{\partial}{\partial t}P(z,t) = 
\frac{\partial^2}{\partial z^2} P(z,t) -
\frac{\partial}{\partial z} P(z,t) \; ,
\end{equation}
where we introduce $z \equiv - \ln T$.
Clearly, the log-normal distribution (\ref{np11}) corresponds to a solution of
Eq.~(\ref{lognorm-eq}) in the absence of any additional boundaries.

In the presence of the sink, which is imposed by the boundary condition
$P(z_{\rm max},t) = 0$ with
\begin{equation}\label{defzmax}
z_{\rm max} = \ln ( \lambda_{\rm max} + 1 ) \simeq \ln \lambda_{\rm max}
\simeq \ln \left(\frac{V^2}{16 c^2} \right) \; ,
\end{equation}
we can straightforwardly find the solution of Eq.~(\ref{lognorm-eq}) by
subtracting from the log-normal distribution (\ref{np11}) a ``mirror''
distribution centered at some $z>z_{\rm max}$ (namely $2 z_{\rm max}+t$)
with a suitable prefactor. This yields the distribution
\begin{eqnarray}
P(z,t) & = & \frac{1}{\sqrt{4 \pi t}} \left[ 
\exp\left( - \frac{(z - t)^2}{4 t} \right) \right. \nonumber \\
& & \left. - e^{z_{\rm max}} \exp\left( - 
\frac{(z - t - 2 z_{\rm max})^2}{4 t} \right) \right]\; , 
\label{lognorm-decay}
\end{eqnarray}
which is defined for $z < z_{\rm max}$.  Clearly, this distribution satisfies
the evolution equation (\ref{lognorm-eq}) as well as the boundary condition
$P(z_{\rm max},t) = 0$ for all $t$ and the initial condition $P(z,0) =
\delta(z)$ for $z < z_{\rm max}$.  

\begin{figure}
\begin{center}
\includegraphics[width=0.95\columnwidth]{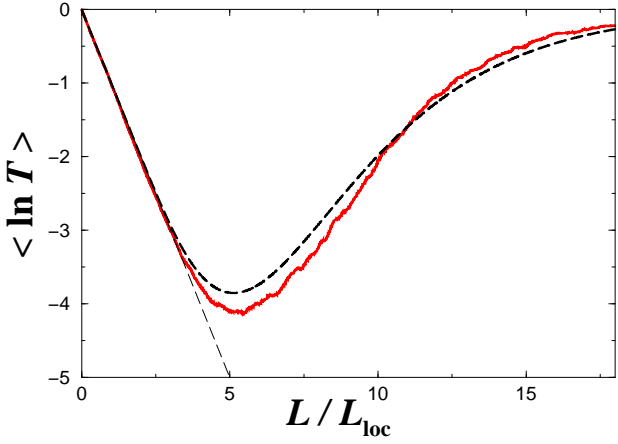}
\end{center}
\caption{(Color online) $\langle\ln T\rangle=-\langle z\rangle$ plotted as a
  function of $t=L/L_{\rm loc}(\kappa)$ in the case of random potentials
  $U_\delta$ characterized by $n_\delta\xi=0.5$ and $\xi/b=\sqrt{2}$. The
  curve is drawn in the case $V/c=450$. The red solid line is the numerical
  result and the black dashed line is the analytical result
  (\ref{zmoy_leak}). The straight (thin dashed) line is the usual DMPK result
  $\langle \ln T\rangle = -t$ [Eq. (\ref{np10})].\label{fig9}}
\end{figure}

The presence of the sink at $z=z_{\rm max}$ explains a phenomenon barely
noticeable in Fig. 7, but exemplified in Fig. \ref{fig9}, namely the departure
of the observed average $\langle \ln T \rangle$ from the usual DMPK result
$\langle \ln T \rangle=-t$. This departure is due to the fact that the
numerically computed average only takes into account the stationary solutions
which --as will be seen from Eq. (\ref{Psurv})-- become less and less
numerous when $t$ increases.  Hence what is computed numerically is the
average of $\ln T=-z$ over the distribution (\ref{lognorm-decay}). This reads
\begin{eqnarray}
\langle z \rangle &=& \int_{-\infty}^{z_{\rm max}}\!\!\!\!
 z \, P(z,t) \, {\rm d}z 
\nonumber \\
&=& \frac{t}{2}\left[
1+ \textrm{erf}\left(\frac{z_{\rm max}-t}{2\sqrt{t}}\right) \right] - 
\nonumber \\
& & e^{z_{\rm max}} 
\left(\frac{t}{2}+ z_{\rm max}\right)\, \textrm{erfc}
\left(\frac{t+z_{\rm max}}{2\sqrt{t}}\right) \; .\label{zmoy_leak}
\end{eqnarray}
where the error function is defined by
\begin{equation}\label{erf}
\mathrm{erf}(x) = \frac{2}{\sqrt{\pi}} \int_0^x \exp(- y^2) \, {\rm d}y \; ,
\end{equation}
and $\textrm{erfc}(x)=1-\textrm{erf}(x)$.

Expression (\ref{zmoy_leak}) is compared in Fig.~\ref{fig9} with the results
of a numerical simulation performed in the case $V/c=450$ (corresponding to
$z_{\rm max}=9.43$) for 10000 random potentials $U_\delta$ of type (\ref{e2a})
characterized by $n_\delta\xi=0.5$ and $\xi/b=\sqrt{2}$ (leading to $L_{\rm
loc}(\kappa)=100\, \xi$). The agreement is seen to be very good. Since the sink
cuts the solutions which are strongly scattered by the random potential, the remaining
stationary states have a higher transmission coefficient. This effect increases with
the sample length $L$, which explains the behavior of the curve in Fig.~\ref{fig9}.

As an other test of the validity of our approach (which amounts to model the
upper boundary $z_{\rm max}$ by a perfect sink and to neglect nonlinear
deformations of the density pattern of the flow close to the threshold) we now
determine the probability for a trajectory to remain below the
boundary. This survival probability reads
\begin{eqnarray}
P_s(t) & = & \int_{-\infty}^{z_{\rm max}} P(z,t) \, {\rm d}z \nonumber \\
& = & \frac{1}{2} \left[1 + \mathrm{erf}
\left(\frac{z_{\rm max} - t}{\sqrt{4 t}}\right) \right] \nonumber \\
& &
- \frac{e^{z_{\rm max}}}{2} \, \mathrm{erfc}
\left(\frac{z_{\rm max} + t}{\sqrt{4 t}}\right) \; .\label{Psurv}
\end{eqnarray}
As anticipated, $P_s(t)$ clearly decreases from $1$ (at $t=0$) to 0 (for large
$t$).  The knowledge of $P_s(t)$ allows to determine the value $L^*$ of the
length of the disordered region beyond which most of the random realizations
lead to a non-stationary flow of the condensate.  We can, most conveniently,
define $L^*$ through the condition
\begin{equation}\label{tstar}
P_s(t^*) = 1/2 \; ,
\end{equation}
with $t^* \equiv L^* / L_{\rm loc}$.
This leads to the implicit equation for the threshold value $t^*$:
\begin{equation}\label{tstar-impl}
\mathrm{erf}\left(\frac{z_{\rm max} - t^*}{\sqrt{4 t^*}} \right) =
e^{z_{\rm max}} \mathrm{erfc}\left( 
\frac{z_{\rm max} + t^*}{\sqrt{4 t^*}} \right) \; .
\end{equation}
This equation can be explicitly solved in the limiting case of large $z_{\rm
  max}$. As it is natural to assume that $t^*$ ought to be of the order of
$z_{\rm max}$, which is the only relevant scale in this equation, we make the
ansatz
\begin{equation}\label{tstar-ansatz}
t^* = z_{\rm max} + \delta t
\end{equation}
and assume (which is to be verified \textit{a posteriori}) that $\delta t$ is
of the order of unity, whereas $z_{\rm max} \gg 1$.  This yields to
lowest non vanishing order
\begin{equation}\label{tstar-left}
\mathrm{erf}\left(\frac{z_{\rm max} - t^*}{\sqrt{4 t^*}}\right) = 
- \frac{\delta t}{\sqrt{\pi z_{\rm max}}} 
\left[ 1 + \mathcal{O}\left(z_{\rm max}^{-1}\right) \right] \; ,
\end{equation}
for the left-hand side of Eq.~(\ref{tstar-impl}) and
\begin{eqnarray}
\lefteqn{e^{z_{\rm max}} \mathrm{erfc}\left( 
\frac{z_{\rm max} + t^*}{\sqrt{4 t^*}} \right) =} \nonumber \\
& = &
e^{z_{\rm max}} \frac{2}{\sqrt{\pi}} 
\int_{\sqrt{z_{\rm max}}}^\infty e^{-y^2}{\rm d}y
\left[ 1 + \mathcal{O}\left(z_{\rm max}^{-2}\right) \right] \nonumber \\
& = & \frac{1}{\sqrt{\pi\, z_{\rm max}}} \left[ 1 + \mathcal{O}\left(z_{\rm
max}^{-1}\right) \right] \; ,
\label{tstar-right}
\end{eqnarray}
for the right-hand side of Eq.~(\ref{tstar-impl}).
This finally results in 
\begin{equation}\label{delta-t}
\delta t = -1 + \mathcal{O}\left(z_{\rm max}^{-1}\right) \; .
\end{equation}
Neglecting terms of the order of $z_{\rm max}^{-1}$, we therefore obtain
for the threshold length
\begin{equation}\label{Lstar}
L^* =  ( z_{\rm max} - 1 ) \, L_{\rm loc} = 
L_{\rm loc} \left[ \ln \left( \frac{V^2}{16 c^2} \right) - 1 \right] \ .
\end{equation}
We emphasize that Eq. (\ref{Lstar}) holds for $z_{\rm max}\gg 1$, i.e., for
$\ln(V^2/16\,c^2)\gg 1$ [see Eq. (\ref{defzmax})]. This is much
more restrictive than the condition $V\gg c$ which is assumed to hold true
when deriving Eqs. (\ref{lognorm-decay}) and (\ref{tstar-impl}).

\begin{figure}
\begin{center}
\includegraphics[width=0.95\columnwidth]{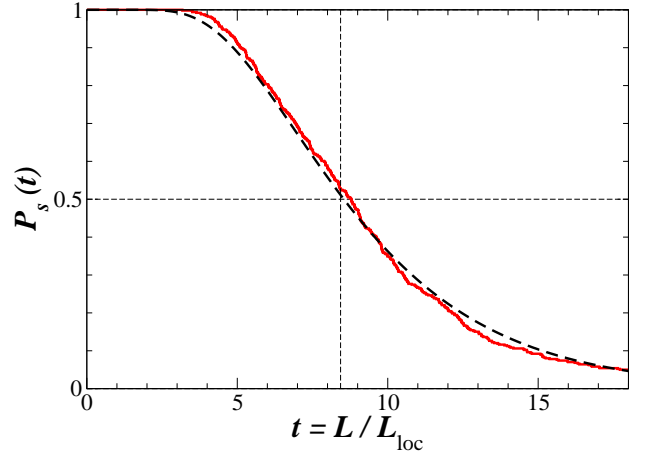}
\end{center}
\caption{(Color online) Fraction of stationary trajectories $P_s(t)$ plotted
  as a function of the length $L$ of the disordered region. The condensate
  flow through disorder potentials $U_\delta$ of type (\ref{e2a}) was
  numerically computed for this purpose (red solid line), at parameters for
  which $V^2/c^2 = 2\times 10^{5}$.  The black dashed line shows the
  analytical prediction of this survival probability $P_s(L/L_{\rm loc})$
  according to Eq.~(\ref{Psurv}). The vertical dashed line marks prediction
  (\ref{Lstar}) for the threshold length $L^*$ at which $P_s(L^*/L_{\rm loc})
  = 1/2$ (horizontal dashed line), namely $L^*/L_{\rm loc} =
  8.433$~.\label{fig10}}
\end{figure}

Figure \ref{fig10} shows a comparison of the analytical predictions
(\ref{Psurv}) and (\ref{Lstar}) with numerical data obtained from the
integration of the time-dependent Gross-Pitaevskii equation (\ref{e1}).  The
condensate flows through a disorder potential $U_\delta$ of type (\ref{e2a})
with $V^2/c^2 = 2\times 10^{5}$.  We see that the fraction of stationary
trajectories $P_s(t)$ is very well described by Eq.~(\ref{Psurv}), and that
the approximate expression (\ref{Lstar}) predicts very well the length $L^*$
at which the crossover length from stationary to time-dependent flow occurs.

\begin{figure}
\begin{center}
\includegraphics[width=0.95\columnwidth]{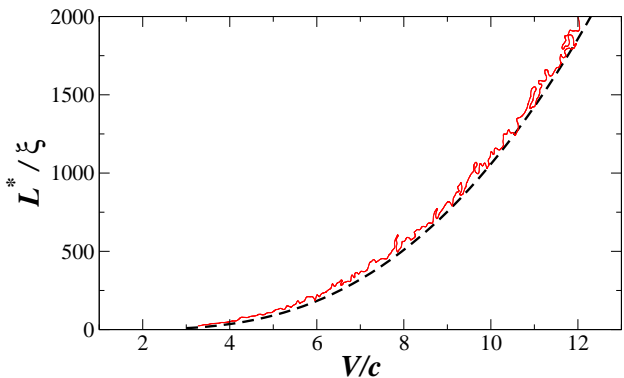}
\end{center}
\caption{(Color online) $L^*$ as a function of $V$ in dimensionless units. The
  black dashed line corresponds to the solution of Eq. (\ref{tstar-impl}) and
  the red solid line corresponds to the value of $L^*$ extracted from numerical
  simulations for a potential $U_\delta$ with the same characteristics as in
  Fig. \ref{fig1} (see the text).\label{fig11}}
\end{figure}

For velocities $V$ not extremely large compared to the speed of sound, the
condition $z_{\rm max}\gg 1$ will not be fulfilled and estimate (\ref{Lstar})
will not be valid, while $\lambda_{\rm max} \gg 1$ might still hold and the
average evolution of the system might still be fairly well described by the
simplified Fokker-Planck equation (\ref{dmpk-large}). In that case, the
implicit equation (\ref{tstar-impl}) has to be solved numerically. In Figure
\ref{fig1} one can see that the numerical solution of Eq.~(\ref{tstar-impl})
(yellow solid line) provides a very reasonable estimate of the boundary
between the bright supersonic region (stationary flows) and the dark
time-dependent region, in a regime of not extremely large $V/c$, where
Eq.~(\ref{Lstar}) fails to properly predict the threshold length $L^*$. The
simulations are performed by solving Eq.~(\ref{e1}) numerically using a
potential of type $U_\delta$ (characterized by $\langle U_\delta\rangle/\mu =
0.025$ and $n_{\delta} \xi = 0.5$). For each $V$ and $L$ we consider 100
realizations of such a potential and statistically determine the quantity
$P_s$, i.e., the fraction of stationary solutions. $P_s$ is plotted in
Fig.~\ref{fig1} using a gray scale (dark, $P_s=0$ ; light blue/gray, $P_s=1$)
as a function of the normalized variables $L/\xi$ and $V/c$ (this
normalization rescales interaction effects). The qualitative agreement of
Figure \ref{fig1} is made quantitative in Figure \ref{fig11}. In this figure
the numerical solution of Eq.~(\ref{tstar-impl}) is compared with its
determination extracted from numerical simulations in the supersonic
regime. More precisely, the solid (red) line in Fig. \ref{fig11} is simply the
contour $P_s=1/2$ in Fig. \ref{fig1}. This corresponds exactly to definition
(\ref{tstar}) of $L^*$. The agreement between the numerical result and the
theory of the present section [dashed curve, solution of
  Eq. (\ref{tstar-impl})] is seen to be excellent \cite{PRL}.

We conclude this section by emphasizing that the existence of an upper
threshold $L^*$ corresponding to lengths of the disordered region beyond which
most of the flows are time-dependent is a genuine nonlinear effect [absent if
  one sets $g=0$ in Eq. (\ref{e1})]. Actually, whereas interactions only
weakly modify the precise value of the localization length, the existence of
the threshold $L^*$ is a remarkable qualitative effect induced by
nonlinearity. Moreover, as illustrated in Figures \ref{fig9} and \ref{fig10},
this effect persists even in the limit $V\gg c$ where naively one would expect
no noticeable consequence of interaction.

\section{Experimental considerations}\label{experiment}

On the basis of the results obtained in the previous section we present here
what are the more favorable experimental configurations for observing Anderson
localization in an interacting Bose-Einstein beam. We also discuss a possible
experimental signature of localization.

\subsection{Appropriate configurations for observing Anderson localization}
\label{apconf}

In the non-interacting regime the only condition for observing Anderson
localization in 1D is that the size of the disordered region should be larger
than the localization length. Then one can observe an exponential decay of
transmission with a log-normal distribution (in the limit $L\gg L_{\rm loc}$).

The situation is more complex when interactions are turned on. What is
particularly interesting is the interplay between localization and
superfluidity. Indeed these two phenomena are conflicting one with the other:
superfluidity is the (counterintuitive) ability to pass over an obstacle
without reflection whereas Anderson localization corresponds to a large
reflection in a domain where one would expect almost perfect transmission. As
a result of the interplay between these two extreme phenomena, and depending
on the fluid velocity and on the sample size, the flow may be stationary and
superfluid, dissipative and time dependent or stationary supersonic (and also
dissipative) \cite{Pau07}. Anderson localization does not occur in the
superfluid region (where the transmission is perfect) and either does not
exist or cannot be clearly identified in the time dependent regime (where
interference effects are washed out \cite{Pau05}); but is truly observed in
the supersonic stationary regime, as demonstrated in Sec. \ref{ssr}.

In that regime, a first experimentally relevant effect is the modification 
of the localization length with respect to its value in the absence
of interactions. This effect is very well described by renormalizing the wave
vector $k$ to $\kappa$ [Eq.~(\ref{spt11})], which means that part of the
kinetic energy available to the flow is taken by interactions. However, as
already discussed in Ref.~\cite{Pau07} this effect is only sizable in a
regime where $V$ is not too large compared to $c$, and is thus relevant
only in the perturbative regime (cf. Fig. \ref{fig5}).

A second experimentally observable effect is the modification of the
localization length due to the correlations of the disordered potential. This
is described by formula (\ref{spt11}) where $\hat{C}$ is the Fourier transform
of the two-point correlation function of the disorder. For the different
potentials considered here, $\hat{C}\equiv 1$ for a potential
$U_\delta$ or is alternatively given by Eqs.~(\ref{e2j}), (\ref{e2k}) and
(\ref{csq}) for correlated potentials. Explicitly this yields
\begin{equation}\label{exp1}
L_{\rm loc}(\kappa)=\frac{\kappa^2}{\sigma} \; ,
\end{equation}
for a potential $U_\delta$ of type (\ref{e2a}); 
\begin{equation}\label{exp2}
L_{\rm loc}(\kappa)=\frac{\kappa^2}{\sigma}\,
\exp\left\{2 \kappa^2\ell_c^2\right\} \; ,
\end{equation}
for a potential $U_{\sss G}$ of type (\ref{e2d});
\begin{equation}\label{exp3}
L_{\rm loc}(\kappa)=\frac{\kappa^2}{\sigma} \,
\exp\left\{2\, \kappa \, \ell_c\right\} \; ,
\end{equation}
for a potential $U_{\sss L}$ of type (\ref{e2d}); and
\begin{equation}\label{exp4}
L_{\rm loc}(\kappa)=\frac{\kappa^2}{\sigma} \, \frac{1}{1-\kappa\, \ell_c} \; .
\end{equation}
for a potential $U_{\sss S}$ of type (\ref{e2l}) (when $\kappa\,\ell_c < 1$).
The validity of these expressions has been tested in Sec. \ref{non-pert}. In
the non-interacting case (i.e., $\kappa=k$), expressions (\ref{exp1}) to
(\ref{exp4}) correspond to a high energy limit and can be obtained through a
first order Born expansion within the phase formalism of
Refs.~\cite{Ant81,Lif88}.  In all three cases, one sees that the localization
length is drastically enhanced due to the non-zero correlation length with
respect to the uncorrelated disorder, Eq.~(\ref{exp1}). In the Gaussian and
the Lorentzian cases the localization length scales exponentially with
$(\kappa \ell_c)^2$ and $\kappa \ell_c$, respectively [see
  Eqs.~(\ref{exp2},\ref{exp3})].  In the case of a speckle potential, the
effect is even stronger: one sometimes speaks of an ``effective mobility
edge'' \cite{Izr99,San07}, meaning that beyond a critical wave-vector (or a
critical velocity) the localization length (\ref{exp4}) is infinite. This is
an artifact of the Antsygina-Pastur-Slyusarev formula (\ref{spt11}) which can
be corrected by going to higher orders (see Refs. \cite{Tes02,Gur09,Lug09}):
the corrections to this result give a localization length which is finite, but
typically larger than any other relevant scale in experimental systems.

Hence, in all the cases the dependence of the localization length with respect
$\kappa$ (i.e., with velocity) is amplified by correlations. Mathematically
this is due to the fact that the denominator in the Antsygina-Pastur-Slyusarev
formula (\ref{spt11}) for the localization length in presence of correlations
tends to zero when $\kappa\ell_c\gg 1$. In order to minimize this effect one
needs to impose the following condition:
\begin{equation}\label{exp5}
  \kappa \, \ell_c \lesssim 1 \quad
\mbox{or} 
 \quad 
V\lesssim V_c=\frac{\hbar}{ m\,\ell_c}=c\, \frac{\xi}{ \ell_c} \; .
\end{equation}
In the r.h.s. of Eq.~(\ref{exp5}) we replaced $\kappa$ by $mV/\hbar$ because
in practice condition (\ref{exp5}) is verified in regimes where $V\gtrsim
3\, c$, i.e., when the approximation $\kappa\simeq k$ is sound.  Note that
this condition is arbitrary and is only superficially analogous to the 3D
Ioffe-Regel criterion \cite{Iof60}. The latter defines a true mobility edge
that separates a metallic from a localized phase whereas Eq.~(\ref{exp5}) only
requires that the localization length does not get too large. Understood in
this sense, the criterion (\ref{exp5}) is exactly equivalent to the definition
of an ``effective mobility edge'' sometimes used in the literature.

\begin{figure}
\begin{center}
\includegraphics[width=0.95\columnwidth]{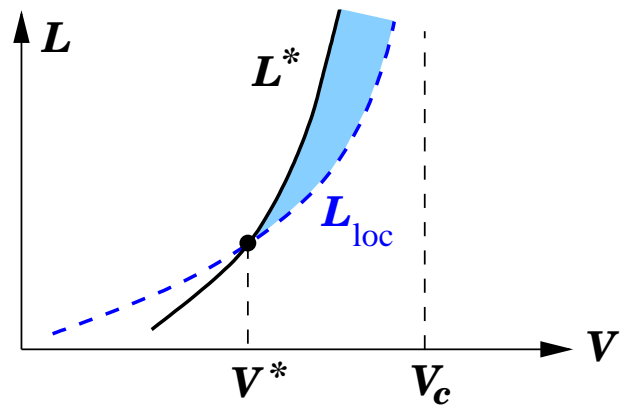}
\end{center}
\caption{(Color online) Schematic phase diagram in arbitrary units. The (blue)
  dashed line is the localization length $L_{\rm loc}$ and the solid line is
  the threshold length $L^*$. $V_c=\hbar/m \ell_c$ is the typical velocity
  beyond which it is almost impossible to observe Anderson localization in a
  realistic system (see text). The (blue) colored zone corresponds to the
  region where Anderson localization can be experimentally observed in
  presence of interaction.\label{fig12}}
\end{figure}

In the absence of interactions it is always possible (at least theoretically)
to define a system with a length $L>L_{\rm{loc}}$ which verifies (\ref{exp5});
i.e., a system where one can observe Anderson localization. If we now turn on
interactions, a major effect is the appearance of a length scale $L^*$
which signals the onset, for $L>L^*$, of a regime of time-dependent flows
(cf. Sec. \ref{sec-lstar}). In this regime, Anderson localization disappears,
and the time-averaged transmission coefficient scales as $1/L$
\cite{Pau05}. This is the most spectacular effect of interactions in the
transport properties of the system.  In order to observe Anderson
localization, the system size should therefore satisfy $L_{\rm loc}< L <
L^*$. In practice, one should be in a regime of parameters such as illustrated
in Fig.~\ref{fig12}: the crossing $L^*>L_{\rm oc}$ has to occur at a velocity
lower than $V_c$.

\begin{figure}
\begin{center}
 \includegraphics[width=0.95\columnwidth]{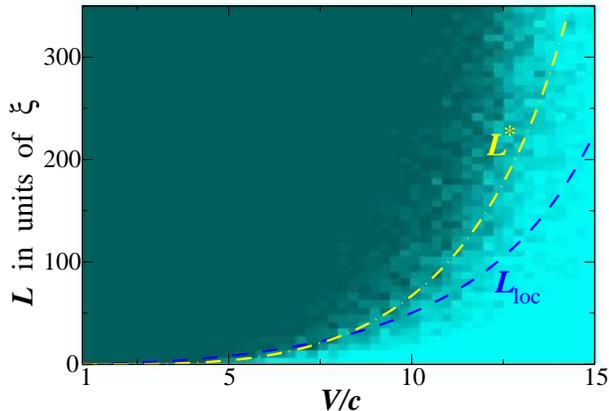}
\end{center}
\caption{(Color online) Phase diagram displaying the fraction of stationary
  trajectory $P_s$ for a beam with velocity $V$ moving in a speckle disorder
  of extent $L$. The figure has been drawn for a potential $U_{\sss S}$ of
  type (\ref{e2l}) characterized by $\ell_c/\xi=0.05$ and
  $\sigma=3.93\,\mu^2\xi$. The light blue region corresponds to a domain of
  stationary flow ($P_s=1$: 100 $\%$ of the solutions are stationary, see the
  explanation in the text); the dark region corresponds to time dependent
  flow. The curves indicating the values of $L_{\rm loc}$ and $L^*$ correspond
  to the analytical results (\ref{exp4}) and to the numerical solution of
  Eq. (\ref{tstar-impl}).\label{fig13}}
\end{figure}

Based on the numerical solution of Eq. (\ref{tstar-impl}) one can show that
the crossing $L^* \ge L_{\rm loc}$ occurs at a velocity $V^*\simeq 7.95\,c$
\cite{795} (see also Fig. \ref{fig1}). This condition only allows the system
to reach a (stationary) regime where $L=L_{\rm loc}$. But if one wants to
observe Anderson localization one should be able to reach a regime where $L^*
\gtrsim L \gtrsim 2 \, L_{\rm loc}$ say, in order to get as close as possible
to the domain of log-normal distribution of transmissions still remaining in
the region of stationary flows. This imposes $V/c\gtrsim 20$. This must be
supplemented by condition (\ref{exp5}), i.e. $V/c\lesssim \xi/\ell_c$.  Hence
the correlation length $\ell_c$ should be smaller or equal to $\xi/20$.
Fig. \ref{fig13} shows the phase diagrams of a one dimensional interacting
beam of condensed atoms moving through a speckle potential in this regime. For
plotting this diagram one has generated 16 random potentials and studied in
each case if a stationary solution exists or not. The dark blue region
corresponds to a domain where no stationary solution exists while the light
blue one corresponds to a domain where all the potentials admit a stationary
solution (the color code is the same as in Fig. \ref{fig1} and is explained in
Sec. \ref{sec-lstar}).  The region between $L_{\rm loc}$ and $L^*$ in Figure
\ref{fig13} is the region where one can observe Anderson localization.

Let us now evaluate the orders of magnitude of the different parameters
allowing one to reach the appropriate regime.  For concreteness we consider a
beam of $^{87}$Rb atoms such as the one of the Atom Optics Group at
Laboratoire Charles Fabry de l'Institut d'Optique. For a correlation length of
0.26 $\mu$m the velocity cut-off for observing Anderson localization is
roughly $V_c\simeq 2.7$ mm/s. Note that in Ref.~\cite{Bil08} the velocity of
the expanding condensate is about 1.6 mm/s, i.e., smaller that $V_c$ as it
should. If we use the parameters of Ref.~\cite{Gue07} ($a\,n_{1}\simeq 0.25$,
$V\simeq 9$ mm/s, $a=5.3$ nm, $n_{1}=45$ atoms/$\mu$m, $c\simeq 0.9$ mm/s and
$\xi\simeq 0.8$ $\mu$m) it is impossible to satisfy condition (\ref{exp5})
because $V\simeq 9$ mm/s $>V_c$ and also because $\ell_c/\xi\simeq
0.3$. However the Atom Optics Group has recently improved the sensitivity of
its detectors which is now close to being able to detect a density as small as
1 atom/$\mu$m. This allows one to work with a smaller density and to improve
the ratio $\ell_c/\xi$ which can be tuned down to the value of $0.05$. Then
the localization length can be selected by tuning the speckle amplitude. For
instance $L_{\rm loc}=0.25$ mm can be obtained by choosing $\langle U_{\sss
  S}\rangle =34$ Hz at $V=1.6$ mm/s. These parameters are close to those used
in Fig. \ref{fig13} and are reachable experimentally. However, for observing
Anderson localization one needs to keep the beam stable for almost 1 s (0.31 s
if we want $L=2L_{\rm loc}$) whereas in the current experiment this is only
possible during 0.1 s, hence it is still a matter of debate to decide if the
observation of Anderson localization of a Bose condensed beam in the presence
of interaction is within the reach of present time technology.

It is also interesting to make a connection between the physics described here
and the recent experiment observation of Anderson localization of a condensate
expanding in a disordered potential performed in the same group
\cite{Bil08}. Contrarily to the propagation of a beam studied in the present
work, Ref. \cite{Bil08} considers the spreading of a wave packet (initially at
rest) in a speckle potential.  After a first stage of expansion, mainly driven
by interactions, the experimental cloud expands with a constant velocity
$V\simeq 1.6$ mm/s but the particle density and the sound velocity are
functions of the position. Therefore it is not possible to place this
experiment on a single point of the phase diagram displayed in
Fig. \ref{fig13}. However one can evaluate the ratio $V/c$ at different
positions, that is at different $L$. For instance, at the typical value
$L=L_{\rm loc}$, Fig. 2 of reference \cite{Bil08} allows to calculate the
sound velocity as well as the healing length $\xi$, yielding the typical
experimental values $V/c\simeq 12$ and $L/\xi\simeq 55$.  Moreover the ratio
between the typical disorder amplitude $\langle U_{\sss S}\rangle$ and the
chemical potential $\mu=g\,n_0$ is the same in Fig. \ref{fig13} and in
experiment: $\langle U_{\sss S}\rangle/\mu=5$ (note however that in the
experimental case this is the value of the {\it local} chemical potential
that matters). Hence, although the experimental setup of Ref. \cite{Bil08}
forbids a direct comparison with the results of the present work, the
estimates of the typical values $V/c\simeq 12$ and $L/\xi\simeq 55$ indeed
locate the experimental system within the regime of Anderson localization of
Fig. \ref{fig13}.

\subsection{Experimental signature}\label{signature}

Once the appropriate regime of parameters for observing Anderson localization
in a Bose condensed beam has been determined, it is also important to identify
possible experimental signatures. In our theoretical approach we use the
transmission coefficient $T$ of the beam over the disordered region as the
relevant parameter. However the measure of $T$ might be experimentally
involved, and we propose here to use an other related quantity, namely the
rate of energy dissipation \cite{Ast04}: $\dot{E}= -V F_{d}$ where
\begin{equation}\label{exp6}
F_d=\int_{\mathbb{R}}{\rm d}x \, n(x,t) \,\frac{\partial U}{\partial x}\; ,
\end{equation}
is the drag force exerted by the beam on the obstacle. Definition
(\ref{exp6}) is quite natural: the force exerted on the obstacle is the mean
value of the operator $\partial_xU$ over the condensate wave function. It is
rigorously justified by the analysis of Ref.~\cite{Pav02} in terms of stress
tensor.  In the stationary case, changing integration from $x$ to $X$ in
(\ref{exp6}), a simple integration by parts yields
\begin{equation}\label{exp10}
F_d=-n_0 \int_{\mathbb{R}} U \, \frac{{\rm d}A^2}{{\rm d}X} \, {\rm d}X 
=n_0 \left(E_{\rm cl}^--E_{\rm cl}^+\right) \; .
\end{equation}
In the r.h.s. of Eq.~(\ref{exp10}) we made use of relations (\ref{e4}) and
(\ref{t1}). 

It has been shown in section \ref{transmis} that $E_{\rm cl}^-=0$
and that under assumption (\ref{t2}) (small nonlinearity and arbitrary
transmission or weak transmission and arbitrary nonlinearity) one has (see
Eqs.~(\ref{t7}) and (\ref{t5})) $E_{\rm cl}^+=2\hbar^2\kappa^2/m(R/T)$ which
yields
\begin{equation}\label{exp11}
F_d=- \frac{2\hbar^2\kappa^2}{m} \, n_0 \, \frac{R}{T} \; .
\end{equation}
In the regime $R\ll 1$, using Eq.~(\ref{spt3}) one recovers from (\ref{exp11})
the perturbative result already obtained in Ref.~\cite{Pav02}: $F_d=-2 n_0 m
|\hat{U}(2\kappa)|^2/\hbar^{2}$ \cite{attention}.

The physics embodied in Eq.~(\ref{exp11}) is rather simple and it is worth
spending some time to discuss it. Consider an incident beam of particles with
density $n_{\rm inc}$ and momentum $p=-\hbar\kappa$ moving from $+\infty$
towards an obstacle at rest. Part of the particles is transmitted (a fraction
$T$) and the other part is reflected (a fraction $R$). The collisions are
elastic and each of the reflected particles experiences an exchange of
momentum $\delta p=2\hbar\kappa$ with the obstacle. During a time $\delta t$
there are $N_{\rm coll}$ collisions and by the law of action and reaction the
obstacle experiences a force
\begin{equation}\label{exp12}
F_d=- N_{\rm coll}\,\frac{\delta p}{\delta t} 
=- \frac{2\hbar^2\kappa^2}{m} \, n_{\rm inc} \, R\; .
\end{equation}
In the r.h.s. of (\ref{exp12}), one has written that $N_{\rm coll}/\delta t$
is the flux of particles colliding with the obstacle i.e.,
$\frac{\hbar\kappa}{m} \, R \, n_{\rm inc}$. Eqs.~(\ref{exp11}) and
(\ref{exp12}) are identical because what we call $n_0$ is the downstream
density of the beam (cf. Fig. \ref{fig2}), i.e., precisely $T\, n_{\rm
  inc}$. Depending on which quantity is held constant ($n_0$ as in the present
paper, or $n_{\rm inc}$) Eq.~(\ref{exp11}) or (\ref{exp12}) is more
appropriate (cf. the discussion of the fixed input and fixed output problem in
Ref. \cite{Pau07b}). This is somewhat reminiscent of the controversy on the
correctness of the Landauer formula (see, e.g., the discussion in
Ref.~\cite{Imr97}).

On the basis of (\ref{exp11}) one sees that the measure of $\dot{E}$ gives
direct informations on the transmission of the interacting beam through the
disordered region, allowing one to reveal in which configuration is the
system. For instance in the localized regime the energy dissipation is high
($\propto 1/T$) and grows exponentially with the size $L$ of the disordered
region, whereas in the perturbative regime $\dot{E}$ is much lower and scales
as $L$.

\section{Conclusion}\label{conclusion}

In the present work we have presented an analysis of the transmission of a
weakly interacting Bose gas incident on a disordered potential.  We have shown
on the basis of numerical and analytical results that there is a regime of
Anderson localization in this system and proposed experimental signature of
this phenomenon. In order to properly identify a ``localized regime'' we have
studied the transmission coefficient and its probability distribution. The
transmission coefficient $T$ is well defined under assumption (\ref{t2}),
which holds in the following regimes : (i) small nonlinearity and arbitrary
transmission or (ii) weak transmission and arbitrary nonlinearity. In other
cases there is no obvious way to define the transmission of the non-linear
beam because one cannot separate in the up-stream region an incident flow from
a reflected one. However, our analysis in terms of $E_{\rm cl}$ and $\lambda$
(defined in Sec. \ref{transmis}) is always valid, even when condition
(\ref{t2}) is not fulfilled. This just means that, out of regime
(\ref{t2}), the connection (\ref{t7}) between $\lambda$ and $T$ is
invalid. But for instance this does not invalidate at all the analysis leading
to the DMPK equation (\ref{dmpk}), and the experimental signature proposed in
Sec. \ref{signature} also remains valid even when it is not possible to
properly define $T$.

We note that the validity of the DMPK approach for non-interacting particles
is a well established fact in the theory of disordered systems. What is
achieved in the present work 
is its extension to the case of interacting particles.  Other studies of
Anderson localization in the presence of interactions have concentrated on the
long time behavior of the time evolution of initial wave packets
\cite{timevolution}. Although those results are still a matter of active
debate in the community, the results of the present work produce strong
evidence of the existence of Anderson localization for weakly interacting Bose
particles (with effective repulsive interaction) propagating through
disordered samples of finite size $L<L^*$.

Although the present study leads to the important conclusion that Anderson
localization in the presence of interaction is possible, it is rather
disappointing to remark that it can be clearly identified only when $V \gtrsim
20 \, c$, i.e., in a regime where interactions do not play a major role (see
the discussion of Sec. \ref{apconf} and also Ref.~\cite{Pau07}). In this
respect, the more interesting and new effect of interactions is the existence
of an upper threshold $L^*$ for the length of the disordered region~: for
$L>L^*$ no stationary flow is possible. As shown in Sec. \ref{sec-lstar},
$L^*$ is directly connected to the probability distribution of the parameter
$\lambda$ and to the localization properties of the system. It would be very
interesting to lead a systematic study of the transmission in the
interaction-induced time dependent regime ($L>L^*$) where the numerical
results of Ref.~\cite{Pau05} indicate a power law decay of the time-averaged
transmission, a signature generally considered of loss of phase coherence and
onset of Ohmic behavior \cite{Dat95,Ber97}. Work in this direction is in
progress.

\begin{acknowledgments}
It is a pleasure to thank B. Altshuler, A. Comtet, J.-L. Pichard and C. Texier
for inspiring discussions. This work was supported by ANR Grants
No. 05--Nano--008--02, No. NT05--2--42103 and No. 08--BLAN--0165--01, by
the IFRAF Institute and by the Alexander von Humboldt Foundation. We
gratefully acknowledge funding by the Excellence Initiative of the German
Research foundation (DFG) through the Heidelberg Graduate School of
Fundamental Physics (Grant No GSC 129/1) and the Global Networks Mobility
Measures the Frontier Innovation Fund of the University of Heidelberg. We
furthermore acknowledge support through the DFG Forschergruppe 760 "Scattering
systems with complex dynamics".
\end{acknowledgments}

\appendix

\section{Distribution of reflection coefficients in the perturbative 
case for a potential of type (\ref{e2d})}\label{app1} 

We give here a demonstration of the perturbative results (\ref{spt4}),
(\ref{spt10}) and (\ref{spt11}) in the special case of a Gaussian disordered
potential $U_g$ of type (\ref{e2d}). A simple way to obtain this results
starts by noticing that any Gaussian noise verifying $\langle
\eta(x)\rangle=0$ and $\langle \eta(x)\eta(x')\rangle=\delta(x-x')$ [and here
  we are specifically interested in $\eta(x)$ that appears in Eq. (\ref{e2d})]
can be written as (see, e.g., Ref.~\cite{texier})
\begin{equation}\label{a1}
  \eta(x)=\lim_{\nu\to\infty}
\frac{1}{\sqrt{\nu}}\sum_{j=-\infty}^{+\infty} \epsilon_j \, \delta(x-X_j)
\; ,
\end{equation}
where $X_j$'s are random positions uniformly distributed on the real axis
with density $\nu$ and mean spacing $1/\nu$ and $\epsilon_j=\pm 1$ is a random
variable (with $\langle \epsilon_j\rangle=0$ and $\langle \epsilon_i
\epsilon_j\rangle=\delta_{i j}$).

In order to calculate the probability distribution of the reflection
coefficient $R$ [whose value is given by (\ref{spt3})] one should first
consider the distribution of
\begin{eqnarray}\label{a2}
  \hat U_g(2\kappa)& = & 
\int_{\mathbb{R}} \!\! {\rm d}x \, U_g(x) \, \ep^{2{\rm i} \kappa x} 
\nonumber \\
& = & \lim_{\nu\to\infty}
\frac{\hbar^2\sqrt{\sigma}}{m}\,
  \frac{\hat w(2\kappa)}{\sqrt{\nu}}\, \sum_{j=0}^{\nu L} 
\epsilon_j\, \ep^{2{\rm i}\kappa X_j} \; .
\end{eqnarray} 
The quantity $\hat U_g(2\kappa)$ as given by (\ref{a2}) is formally equivalent
to the position $z$ of a particle performing a random walk in the complex
plane after $N=\nu L$ iterations.  The particle is initially at the origin and
performs jumps of constant amplitude
$s=\frac{\hbar^2}{m}|\hat{w}(2\kappa)|\sqrt{\sigma/\nu}$ with random
direction. Denoting by ${\rm d}^2P=p(z,N){\rm d}x{\rm d}y$ the probability to
find the particle in the domain ${\rm d}x{\rm d}y$ around $z$ after $N$ steps,
if $N\gg 1$ [which is ensured by taking the limit $\nu\to\infty$ in
  (\ref{a2})], the central limit theorem yields
\begin{equation}\label{a3}
p(z,N)=\frac{1}{\pi\, N\, s^2} 
\exp\left(-\frac{|z|^2}{N\, s^2}\right) \; .
\end{equation}
It is then a simple exercise to get the 
distribution of $|z|^2$. One obtains
\begin{equation}\label{a4}
P(|z|^2,N)=
\frac{1}{\langle |z|^2\rangle}
\exp\left[ -\frac{|z|^2}{\langle |z|^2\rangle}\right] \; ,
\end{equation}
where
\begin{equation}\label{a5}
\langle |z|^2\rangle =s^2\,N=
\sigma (\hbar^2/m)^2 |\hat{w}(2\kappa)|^2\,L\; .
\end{equation}
From relation (\ref{spt3}), Eq.~(\ref{a5}) immediately yields
the announced probability distribution (\ref{spt4}) with
\begin{equation}\label{a6}
\langle R \, \rangle = \frac{m^2}{\hbar^4\,\kappa^2}\langle |z|^2\rangle
=\frac{\sigma\, L}{\kappa^2} \, |\hat{w}(2\kappa)|^2 \; .
\end{equation}
For a potential $U_g$ of type (\ref{e2d}), $|\hat{w}|^2=\hat{C}$ [see
  Eq.~(\ref{e2g})] and Eq.~(\ref{a6}) demonstrates in this case the validity
of Eqs.~(\ref{spt10}) and (\ref{spt11}).

\section{Derivation of the DMPK equation (\ref{dmpk})}\label{app2} 

In this appendix, we explain how to obtain the DMPK equation (\ref{dmpk})
starting from the discrete Langevin equation (\ref{np6}).

Let us consider a generic situation where $\lambda_n$ obeys a
stochastic recursion relation of the type
\begin{equation}\label{b1}
\lambda_{n+1}-\lambda_n=F(\lambda_n,\zeta_n) \; ,\end{equation}
with uncorrelated random variables $\zeta_n$~: 
\begin{equation}\label{b2}
\langle\zeta_{n_1}\zeta_{n_2} \cdots \,
\zeta_{n_N}\rangle=C_N \; \delta_{n_1 n_2} \cdots \, \delta_{n_1 n_N} \; .
\end{equation}
It is clear that under assumption (\ref{gnp2bis}), Eq.~(\ref{np6}) is of
type (\ref{b1}) with all the odd $N$ averages in (\ref{b2}) being zero and
$C_2=1/2$ [cf. Eq.~(\ref{varxi})].

Let $P(\lambda,n){\rm d}\lambda$ be the probability that $\lambda_n$ lies in
the interval $\lambda$, $\lambda+{\rm d}\lambda$. One can express
$P(\lambda,n)$ as
\begin{equation}\label{b3}
P(\lambda,n)=\langle \delta(\lambda_n-\lambda)\rangle = \left\langle 
\int_\mathbb{R} \frac{{\rm d}k}{2\pi}
\, \ep^{{\rm i} k (\lambda_n-\lambda)}
\right\rangle\; .
\end{equation}
This yields
\begin{eqnarray}\label{b4}
& & P(\lambda,n+1)-P(\lambda,n) =  \nonumber \\
& & \nonumber \\
& & \displaystyle\left\langle 
\int_\mathbb{R} \frac{{\rm d}k}{2\pi}
\, \ep^{{\rm i} k (\lambda_n-\lambda)}
\left(\ep^{{\rm i}kF(\lambda_n,\zeta_n)}-1\right)\right\rangle = \nonumber \\
& &  \nonumber \\
& & \displaystyle\sum_{\ell=1}^\infty \frac{(-1)^\ell}{\ell\, !}\,
\frac{\partial^\ell}{\partial\lambda^\ell}
\left\langle F^\ell(\lambda_n,\zeta_n) 
\int_\mathbb{R} \frac{{\rm d}k}{2\pi}
\, \ep^{{\rm i} k (\lambda_n-\lambda)} \right\rangle = \nonumber \\
& & \nonumber \\
& & \displaystyle\sum_{\ell=1}^\infty \frac{(-1)^\ell}{\ell\, !}\,
\frac{\partial^\ell}{\partial\lambda^\ell}
\left\langle F^\ell(\lambda,\zeta_n) \delta(\lambda_n-\lambda)\right\rangle
\; . 
\end{eqnarray}
Using the fact that $\lambda_n$ depends on the variables $\zeta_1$, $\zeta_2$
\ldots $\zeta_{n-1}$ but not on $\zeta_n$ (as can be seen directly from
(\ref{b1})) one can write the last of Eqs.~(\ref{b4}) as
\begin{eqnarray}\label{b5}
& & 
P(\lambda,n+1)-P(\lambda,n) = \nonumber \\
& & \nonumber \\
& & \displaystyle\sum_{\ell=1}^\infty \frac{(-1)^\ell}{\ell\, !}\,
\frac{\partial^\ell}{\partial\lambda^\ell}
\left\{\left\langle F^\ell(\lambda,\zeta_n)\right\rangle P(\lambda,n)\right\}
\; .
\end{eqnarray}
In the case of Eq.~(\ref{np6}) one has $F(\lambda,\zeta)= (1+2\lambda)/
\kappa^2b^2+ 2(\lambda^2+\lambda)^{1/2}\zeta/\kappa b$ and the successive
moments of $F$ read
\begin{equation}\label{b6}
\langle F(\lambda,\zeta_n)\rangle=\frac{1+2\lambda}{\kappa^2 b^2} \; ,
\end{equation}
\begin{equation}\label{b7}
\langle F^2(\lambda,\zeta_n)\rangle=\frac{2(\lambda^2+\lambda)}{\kappa^2b^2}
+{\cal O}\left(\frac{1}{\kappa^4b^4}\right) \; ,
\end{equation}
with all the other moments being of order $1/(\kappa^3 b^3)$ or more, i.e.,
negligible in regime (\ref{np4a}). Eq.~(\ref{b5}) thus reads
\begin{eqnarray}\label{b8}
& & \displaystyle \kappa^2b^2\left[P(\lambda,n+1)-P(\lambda,n)\right] = 
\nonumber \\
& & \nonumber \\
& & \displaystyle
-\frac{\partial}{\partial\lambda}\left[(1+2\lambda)P\right] + 
\frac{\partial^2}{\partial\lambda^2}\left[(\lambda^2+\lambda)P\right] = 
\nonumber \\
& & \nonumber \\
 & & \displaystyle
\frac{\partial}{\partial\lambda}
\left[\lambda(\lambda+1)\frac{\partial P}{\partial\lambda}\right]
  \; .
\end{eqnarray}
In the continuous limit, defining $t=n/(\kappa^2 b^2)$, the l.h.s. of
Eq.~(\ref{b8}) is simply the first derivative of $P$ with respect to $t$ and
(\ref{b8}) reduces to Eq.~(\ref{dmpk}) of the main text.


\begin{thebibliography}{99}
\bibitem{rem} Note in particular that condition (\ref{t2}) is always
  fulfilled in absence of interaction since in this case the speed of sound is
  $c\equiv 0$.
\bibitem{rem_speckle} Eq.~(\ref{e2l}) corresponds to a typical physical
  situation where the potential seen by the particles
is proportional to the intensity of an electric field whose
  components are random Gaussian variables.
\bibitem{ohm} In Ref. \cite{Pau07} this region has been improperly called
  ``ohmic''.
\bibitem{rem7} In order to complete the proof of this equivalence one has to
  show that the mean value $\langle \lambda\rangle = t=L/L_{\rm loc}(\kappa)$
  obtained from (\ref{dmpk0}) for a sample of length $X=L$ is identical to the
  result (\ref{spt5}). This amounts to show that for a potential of the form
  (\ref{e2a}) one has $\langle |\hat{U}_\delta(2\kappa)|^2\rangle =
  (\hbar^2/m)^2\sigma L$, which is easily obtained.
\bibitem{PRL} Note that in Ref. \cite{Pau07} the theoretical determination of
  $L^*$ was obtained using a criterion different from (\ref{tstar}) resulting
  in a less precise estimate at high $V/c$.
\bibitem{795} Note that the value $V^*/c=7.95$ does not depend on the type of
  correlation of the disordered potential considered since it is simply the
  value of $V/c$ for which Eq. (\ref{tstar-impl}) admits the solution $t^*=1$.
\bibitem{attention} Note that there is a factor of 2 difference between the
  present definition (\ref{pt4}) of $\kappa$ and the one used in
  Ref.~\cite{Pav02}.
\end{thebibliography}
\end{document}